\documentclass[sigconf]{acmart}

\AtBeginDocument{%
  }

\copyrightyear{2026}
\acmYear{2026}
\setcopyright{cc}
\setcctype{by}
\acmConference[CHI '26]{Proceedings of the 2026 CHI Conference on Human Factors in Computing Systems}{April 13--17, 2026}{Barcelona, Spain}
\acmBooktitle{Proceedings of the 2026 CHI Conference on Human Factors in Computing Systems (CHI '26), April 13--17, 2026, Barcelona, Spain}
\acmPrice{}
\acmDOI{10.1145/3772318.3791159}
\acmISBN{979-8-4007-2278-3/2026/04}

\usepackage[utf8]{inputenc} 
\usepackage{csquotes}
\usepackage{graphicx}
\usepackage{subcaption}
\usepackage{soul}

\begin{document}

\title{PlayWrite: A Multimodal System for AI Supported Narrative Co-Authoring Through Play in XR}

\author{Esen K. Tütüncü}
\email{esenkucuktutuncu@ub.edu}
\orcid{1234-5678-9012}
\affiliation{%
  \institution{Institute of Neurosciences of the University of Barcelona}
  \city{Barcelona}
  \country{Spain}
}

\author{Qian Zhou}
\email{qian.zhou@autodesk.com}
\affiliation{%
  \institution{Autodesk Research}
  \city{San Francisco}
  \country{USA}}

\author{Frederik Brudy}
\email{frederik.brudy@autodesk.com}
\affiliation{%
  \institution{Autodesk Research}
  \city{Toronto}
  \country{Canada}
}

\author{George Fitzmaurice}
\email{george.fitzmaurice@autodesk.com}
\affiliation{%
 \institution{Autodesk Research}
 \city{Toronto}
 \country{Canada}}

\author{Fraser Anderson}
\email{fraser.anderson@autodesk.com}
\affiliation{%
 \institution{Autodesk Research}
 \city{Toronto}
 \country{Canada}}

\renewcommand{\shortauthors}{K. Tütüncü et al.}
\begin{abstract}
Current AI writing tools, which rely on text prompts, poorly support the spatial and interactive nature of storytelling where ideas emerge from direct manipulation and play. We present \textit{PlayWrite}, a mixed-reality system where users author stories by directly manipulating virtual characters and props. A multi-agent AI pipeline interprets these actions into \textit{Intent Frames}—structured narrative beats visualized as rearrangeable story marbles on a timeline. A large language model then transforms the user's assembled sequence into a final narrative. A user study (N=13) with writers from varying domains found that PlayWrite fosters a highly improvisational and playful process. Users treated the AI as a collaborative partner, using its unexpected responses to spark new ideas and overcome creative blocks. \textit{PlayWrite} demonstrates an approach for co-creative systems that move beyond text to embrace direct manipulation and play as core interaction modalities.
\end{abstract}



\begin{teaserfigure}
  \includegraphics[width=\textwidth]{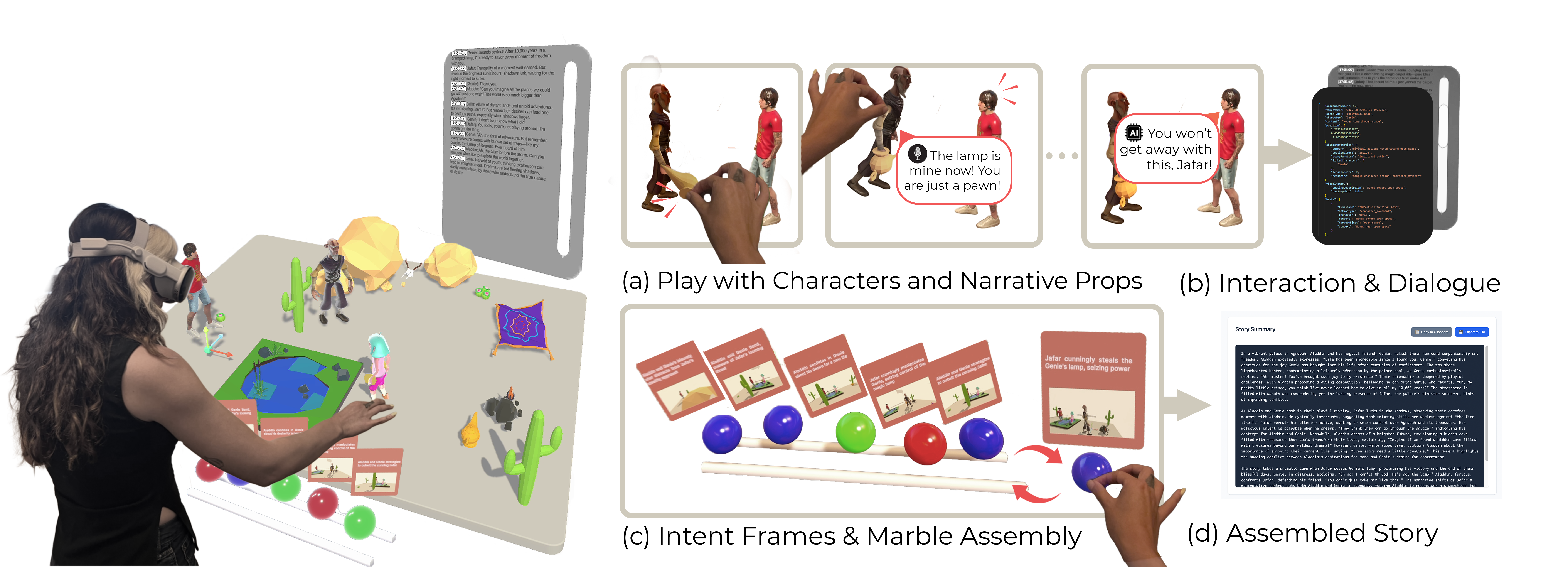}
  \Description{Overview of the system with a user wearing a mixed-reality headset standing at a tabletop surface. Small 3D characters and props such as Aladdin, a carpet, and some rocks are positioned on the table. A speech bubble and icons suggest that gestures and speech are observed by the system}
  \caption{Overview of \textit{PlayWrite}. Users create stories by 
  (a) playing with characters and narrative props, 
  (b) engaging in real-time dialogue with AI-driven characters, 
  (c) assembling intent frames and marbles into narrative structure, 
  and (d) exporting an assembled story. 
  The system situates authoring in playful multimodal interaction, combining spatial manipulation and dialogue with AI.}
  \label{fig:teaser}
\end{teaserfigure}


\maketitle

\section{Introduction}

Stories are among humanity’s oldest tools for meaning-making. From myths enacted around fires to contemporary cinema, storytelling has always been more than linguistic expression: it is also physical, spatial, and social. We tell stories by gesturing with our hands, moving bodies across stages, or arranging toys on a floor. These embodied practices highlight a fundamental truth: creativity often emerges through improvisation in space, not solely through words.

Recent advances in large language models (LLMs) have enabled machines to generate expressive narratives, capable of mimicking different tones, genres, and voices. Yet, the dominant interaction paradigm remains constrained to text input. Typing prompts may suffice for certain tasks, but they poorly support the improvisational and embodied nature of how people often create, especially in creative writing, performance, education, and game design \cite{lakoff1999review,dourish2001action,sawyer2024explaining}. Writers and storytellers frequently develop ideas through sketches, blockings, or play. For many creators outside of professional writing such as hobbyists, performers, or educators, traditional text interfaces can inhibit rather than enhance imagination.

While prior work in narrative generation has focused on algorithmic story structures or natural language co-authorship, less attention has been given to how embodied interaction might shape generative storytelling. In particular, few systems allow users to direct dialogue, character action, and scene composition through embodied input, nor do they support iterative editing of AI-generated narratives through visual, manipulable interfaces \cite{10.1145/3698061.3726933, zhang2022storybuddy, clark2018creative}.


In this paper, we introduce \textbf{PlayWrite}, a proof-of-concept mixed reality (XR) system for AI-assisted storytelling. 
Instead of typing prompts, users stage their stories in XR: manipulating virtual objects, positioning characters, and enacting scenes within their physical environment. The system treats gestures and spatial arrangements as meaningful narrative input, transforming them into structured representations such as editable character dialogue and story beats. These are distilled into what we call \textit{Intent Frames} and visualized as modular \textit{Story Marbles} along a timeline that users can rearrange or refine. 
Finally, PlayWrite composes story drafts using an LLM, enabling authorship through play rather than prose.  PlayWrite is designed for writers and storytellers who develop ideas through performance, improvisation, and spatial reasoning rather than typing alone. Our goal is not to replace traditional writing, but to offer a complementary pathway for creators who think visually and spatially. For these users, play and embodied experimentation can surface character intentions, emotional dynamics, and narrative alternatives that may not emerge through text-first interfaces.  The contribution of PlayWrite lies not in producing polished stories, but in supporting creative exploration through embodied co-creation with AI.

Our contributions are threefold:
\begin{itemize}
    \item A \textbf{system} for AI-assisted storytelling driven by play, direct manipulation, and speech interaction.
    \item The concept of \textbf{Intent Frames}, higher-level abstractions that condense sequences of user inputs into semantic, AI-readable interpretations, which serve as building blocks for narrative arcs.
    \item A \textbf{multi-agent architecture} (Environment Agent, Social Agent, Narrator Agent, and Intent Frame Agent) that collectively interprets embodied performance to generate narrative outputs.
\end{itemize}

Rather than relying solely on textual commands, PlayWrite explores how physical play can serve as a complementary mode of human--AI co-creativity. We situate our work at the intersection of generative storytelling, tangible and embodied interaction, and creativity support tools, offering a system that privileges intuition, improvisation, and spatial engagement.

\section{Related Work}

Our research builds on four areas at the intersection of HCI and AI: (1) play and puppetry as creative expression, (2) AI-assisted narrative generation, (3) embodied and multimodal interaction, and (4) user agency in co-creative systems. Together, these domains illuminate the challenges of designing narrative technologies that move beyond text to embrace embodied, improvisational, and agentic forms of interaction.

\subsection{What is Play?}

Play has long been recognized as a central mode of human expression and meaning-making. 
Huizinga famously defined play as ``a voluntary activity or occupation, executed within fixed limits of time and place, according to rules freely accepted but absolutely binding, having its aim in itself and accompanied by a feeling of tension, joy, and the awareness that it is different from ordinary life''~\cite{huizinga1971homo}. 
This notion of the ``magic circle''  (a temporary, physical or conceptual space set apart from the real world for play) emphasizes how play is bounded, distinct from ordinary activity, and yet deeply generative of meaning.
Building on this foundation, Sicart's \textit{Play Matters} expands play beyond rules and boundaries to highlight its contextual, expressive, and transformative qualities~\cite{sicart2014play}. Sicart characterizes play as \emph{contextual}, unfolding within a tangled world of people, things, and spaces; \emph{autotelic}, with purpose emerging from the interaction itself rather than external outcomes; \emph{creative}, affording improvisation and divergent meaning-making; and \emph{carnivalesque}, balancing creation and destruction with joy and surprise.


HCI work has also described creative writing with AI as a form of play, where the tool functions less as a deterministic author and more as a partner in exploration~\cite{clark2018creative, chakrabarty2024creativity}. 
Together, these perspectives frame play not merely as entertainment but as a design lens for interactive systems, situating our approach to narrative authoring in a lineage of playful, improvisational practice.

\subsection{Narrative Generation}
 Theatrical research provides an important foundation for understanding narrative as something discovered through embodied performance rather than authored exclusively through text. Laurel’s \textit{Computers as Theater}\cite{laurel2013computers} frames interaction as dramatic enactment, while puppetry scholarship \cite{tillis1992toward, kaplin1999puppet, posner2014routledge} conceptualizes meaning as emerging from the negotiation between performer and object. This perspective highlights the potential of play, improvisation, and physical manipulation as modes of narrative construction.

Computational approaches to narrative generation have a long history in both AI and HCI. 
Early work focused on linear story generation through automated screenwriting and planning-based systems such as \textit{TALE-SPIN}~\cite{meehan1977tale}, \textit{Universe}~\cite{lebowitz1985story}, followed by real-time interactive drama experiences such as \textit{Façade}~\cite{mateas2003faccade} which operationalized the principles of drama management introduced by the \textit{Oz Project}~\cite{weyhrauch1997guiding}. 

Subsequent work extended on this, such as the narrative planners balancing plot coherence and character autonomy~\cite{riedl2010narrative}, narrative mediation frameworks~\cite{winslade2000narrative}, and surveys framing interactive narrative as an AI systems challenge~\cite{riedl2013interactive} exemplify this shift. 
Co-authoring tools such as \textit{Say Anything}~\cite{swanson2006say}, and crowd-sourced story graphs~\cite{li2013story} explored open-ended authorship.

Finally, contemporary projects move beyond text toward visual or spatial authoring. 
Systems such as \textit{DirectGPT}~\cite{masson2024directgpt}, \textit{TaleStream}~\cite{chou2023talestream}, and \textit{Patchview}~\cite{chung2024patchview} enable users to manipulate narrative units directly on a canvas or timeline.  Similarly, Visual story-writing \cite{10.1145/3746059.3747758} used visual representations of story elements to support writing and revising narrative text. \textit{Toyteller}~\cite{chung2025toyteller} extended this trend by framing narrative assembly as toy-like manipulation, demonstrating the potential of play as a generative metaphor. 

Yet these systems remain largely symbolic, limited in their contextualization of space and multimodal input. 
In contrast, \textit{PlayWrite} leverages play not only metaphorically but also materially, using direct manipulation, voice, and spatial relations as drivers of narrative intent.

\subsection{Multimodal Storytelling}

 Beyond text-based approaches, a growing body of work has investigated multimodal narrative authoring. These systems recognize that storytelling often involves gestures, visual arrangements, and performance in addition to words. Early work on multimodal interfaces demonstrated how speech and gesture could be combined to construct shared spatial references, most notably Bolt’s “Put That There”~\cite{10.1145/800250.807503}.
More recent systems such as  \textit{Narratron}~\cite{zhao2023narratron} and \textit{Story3D-Agent}~\cite{huang2024story3d} integrate speech with visual interaction to generate stories that combine language and spatial representation. 

Puppetry-inspired systems similarly explore how tangible manipulation can shape character expression and narrative development. 
Projects using physical puppets, digital avatars, or motion-tracked performance illustrate how narrative can emerge through bodily manipulation and staged interaction~\cite{kim2018enhancing,echeverri2024constructing, jacob2013viewpoints}. 

Visual and spatial authoring tools expand these modalities further. 
\textit{Novella}~\cite{green2018novella} and \textit{CodeToon}~\cite{suh2022codetoon} embed agency into visual layouts, while systems such as \textit{Paratrouper}~\cite{10.1145/3706598.3714242} and \textit{Narrative Motion Blocks}~\cite{10.1145/3715336.3735766} introduce novel metaphors for casting characters or composing animation sequences. Other tools such as \textit{WhatIF}~\cite{10.1145/3698061.3726933} and \textit{WhatELSE}~\cite{10.1145/3706598.3713363} scaffold reflective decision-making by visualizing narrative paths and counterfactuals. 

The continuous development of XR platforms has further expanded these possibilities, making multimodal interaction—through gesture, proxemics, and voice—a natural part of the authoring process. 
While dedicated tools for XR-based storytelling remain limited \cite{geigel2020digital}, the affordances of mixed reality environments highlight the potential for users to freely express narrative intent in space, staging characters and actions through the same modalities they use in everyday interaction. By situating narrative generation within XR, \textit{PlayWrite} builds on this potential, positioning multimodal input not as an auxiliary feature but as the core means of shaping stories.

\subsection{User Agency and Control in AI Systems}


A longstanding challenge in co-creative systems is balancing user agency with AI autonomy. Mixed-initiative interfaces have explored ways for both humans and AI to contribute meaningfully while ensuring users retain ownership~\cite{horvitz1999principles,riedl2013interactive}, and creativity support tools have emphasized transparency—enabling users to see how outputs are produced and steer them when needed~\cite{shneiderman2007creativity,amershi2014power}. Recent work shows that AI suggestions can shape topic selection~\cite{poddar2023ai}, alter user opinions~\cite{jakesch2023co}, and affect cognitive framing~\cite{bhat2023interacting}, while prompting strategies mediate users' sense of authorship~\cite{dang2023choice}. These findings underscore that narrative generation systems must balance coherence with responsiveness to user intention.
Other approaches emphasize modular or scaffolded interaction, such as \textit{AI Chains}, which decompose generation into editable steps~\cite{wu2022ai}, or \textit{Clarify}, which supports natural language corrections for iterative refinement~\cite{lee2024clarify}. Frameworks such as Haase and Pokutta’s theory of human–AI authorship~\cite{haase2024human} and surveys of drama management and emergent storytelling~\cite{roberts2008survey,trichopoulos2023survey} reinforce the importance of balancing system autonomy with user control.

Together, these studies underscore the importance of designing co-creative systems that are both flexible and interpretable. 
In the context of storytelling, this means that users must be able to guide narrative direction, understand how their actions are interpreted, and intervene when necessary without being forced to micromanage every detail.

\label{design}
\section{Motivation and Design Goals}

Play is a fundamental part of our lives, offering a rich avenue for exploring ideas through action and experimentation \cite{sicart2014play}. It stimulates and facilitates creativity \cite{mainemelis2006ideas}, allowing us to explore, challenge, and create. Through play, we engage in creative expression, constructing objects, rules, and spaces in dynamic and meaningful ways. This process fosters a shared yet deeply personal form of expression \cite{sicart2014play}. 
Even small gestures during play, such as moving two dolls closer together, can carry significant social and narrative meaning \cite{chung2025toyteller}. 
These interactions often encapsulate rich social contexts and individual emotions. Moreover, play is inherently enjoyable and engaging. It brings joy and reduces perceived effort \cite{quarrington2020role}. It has value beyond its utility, bringing moments of relaxation and fun. Playful systems often naturally invite curiosity \cite{lucero2014playful}, motivating our engagement with a positive and creative mindset.

The creative and joyful aspects of play make it a promising foundation for AI-powered storytelling. By leveraging the spontaneous and expressive qualities of play, AI systems can capture nuanced authorial intents, allowing for the creation of dynamic stories that naturally emerge from user interactions and improvising dialogue. While most AI-powered story authoring systems focus on typed input, Toyteller \cite{chung2025toyteller} explored symbolic play to create stories using text and 2D symbol motions, demonstrating the promise of integrating play into AI-powered storytelling. We extend this concept by exploring how spatial play with spontaneous dialogue can enhance the storytelling experience. By incorporating spatial movements with real-time voice interactions, the system enables the user to become an active co-creator of the story and shape the story through gestural interactions and improvised dialogue.
Inspired by prior work in play, AI-powered story authoring, and multimodal storytelling, we developed the following three design goals to guide the design of our system:

\begin{description}
  \item[\textbf{DG1}] \textbf{Provide a playful experience that invites curiosity and improvisation.}
    Interactions should go beyond utility and cultivate a sense of open-ended play. The system should encourage users to try out new ideas without fear of failure, explore alternative story directions, and remain receptive to surprising or unpredictable outcomes. By supporting spontaneous dialogue, playful staging, and emergent dynamics, the system can keep users engaged and curious about what might happen next, while also fostering joy and improvisation in the process.

  \item[\textbf{DG2}] \textbf{Enable expressive prompting using multimodal input.}
    The system should support multiple channels of expression---including speech, direct manipulation, and spatial arrangement---that complement one another while allowing the users to choose the modality that best matches their creative intent at a given moment, whether it is narrating dialogue, physically staging an action, or positioning characters in space. 

  \item[\textbf{DG3}] \textbf{Translate spatial manipulations into narratively meaningful events that preserve intent.}
    Spatial interactions such as moving a character, grabbing an object, or changing spatial relationships should be interpreted as more than surface-level actions. The system should capture the user’s underlying narrative intent and translate these manipulations into story events. At the same time, the AI should act as a collaborator that extends and elaborates on these contributions while maintaining the user’s direction, offering variations or alternative continuations that improve the narrative. 

\end{description}

\section{PlayWrite}

\textit{PlayWrite} is a story authoring system that enables users to craft stories through improvising dialogue and direct manipulations in XR  Fig \ref{fig:teaser}). 
Rather than composing narratives purely in text, users stage and direct characters in a mixed-reality environment  (DG1), where gestures, dialogue, and object manipulations are interpreted as narrative intent  (DG2). 
The system transforms these multimodal interactions into structured narrative units, allowing stories to naturally emerge from user interactions  (DG3). 

\subsection{XR Interface}

\begin{figure*}[h]
    \centering
    \includegraphics[width=0.98\linewidth]{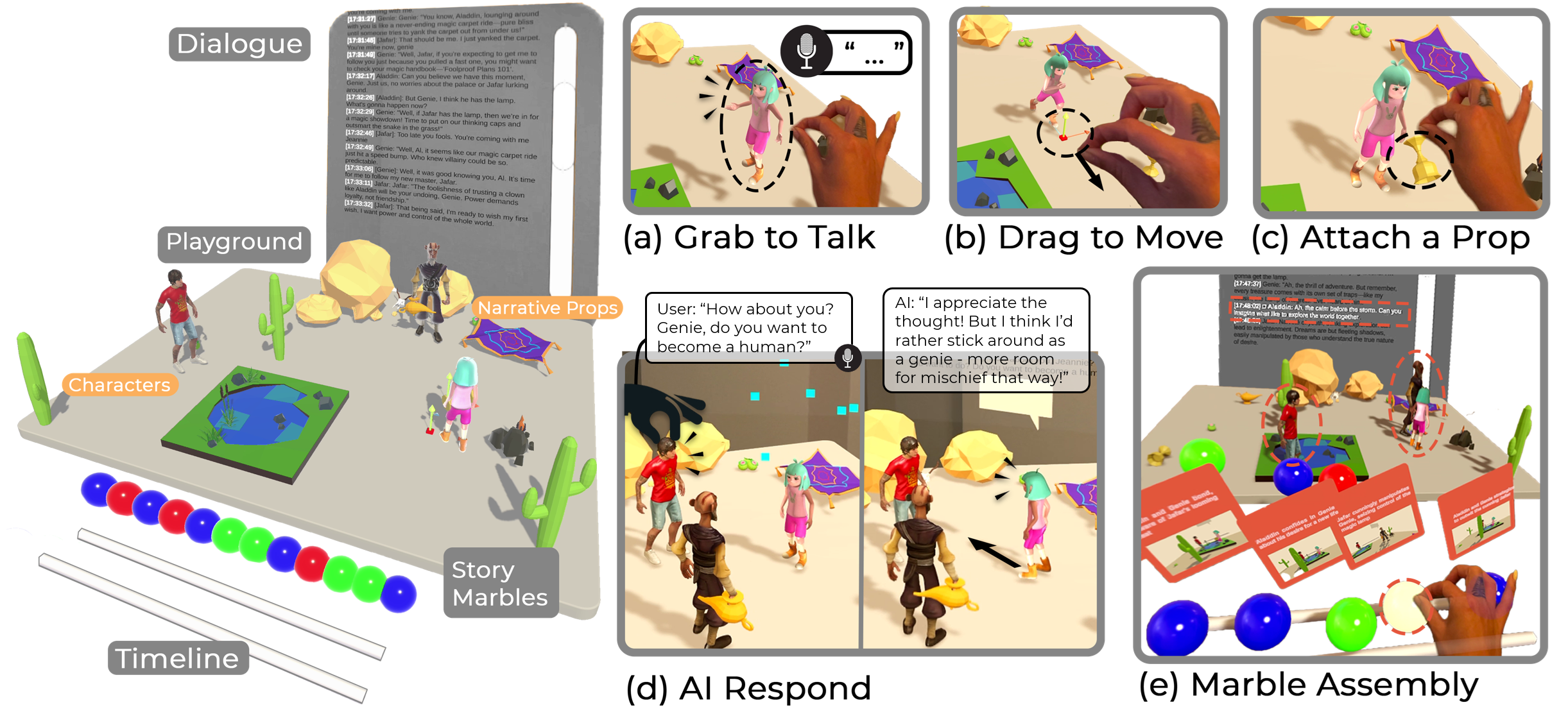}
    \Description{Different views of the PlayWrite interface layout. The main 3D playground shows characters and props. A dialogue window appears above characters when they speak. Below, a horizontal timeline displays circular “story marbles” that can be selected and rearranged.}
    \caption{PlayWrite Interface. (a) \textbf{Grab to Talk:} picking up a character triggers user's voice input. 
  (b) \textbf{Drag to Move:} repositioning the character's transform handle for moving them in space with character animations. 
  (c) \textbf{Attach a Prop:} objects can be linked to characters to contextualize actions. 
  (d)  \textbf{AI Respond:} the system generates dialogue and narrative continuation in response to user actions. 
  (e) \textbf{Marble Assembly:} users arrange story marbles on a timeline, each representing an Intent Frame, to sequence and refine the story. 
  Dialogue and character locations are replayed when a marble is selected, supporting reflection and revisiting.}
    \label{fig:interface}
\end{figure*}

We designed the \textit{PlayWrite} interface in passthrough XR, layering narrative elements directly into the user’s physical space (Fig \ref{fig:interface}). The interface is composed of four main parts:  
1) a \textit{story playground}, a bounded platform where characters and narrative props are placed in front of the user;  
2) a \textit{dialogue history panel}, which records and displays the evolving conversation, allowing users to review or revisit past exchanges;  
3) an \textit{intent frame view}, where system-generated narrative abstractions appear as cards and marbles; and  
4) an \textit{assembly space}, where users organize these marbles into the stories (Figure  \ref{fig:interface}e).

Users are able to interact with characters and narrative props through a combination of direct manipulation and spontaneous dialogue. 


\subsubsection{Direct Manipulation}
\label{Direct Manipulation}

 PlayWrite supports embodied storytelling through spatial manipulation of characters and props (DG2). Each character in the scene has two mutually exclusive interaction areas: a movement handle and a dialogue trigger area. These areas ensure unambiguous intent recognition between locomotion and speech.

\textbf{Grab to Talk.}
Users can grab the body of a character to talk (Figure~\ref{fig:interface}a), which initiates dialogue. This triggers an idle-to-talking animation, activates the speech subsystem, and hides the transform handle to prevent accidental gesture conflict. A short delay ensures animations start before voice input is captured, enabling fluid transition from grab to speech.

\textbf{Drag to Move.}
Movement is controlled via grabbing a transform handle located near each character (Figure~\ref{fig:interface}b). These handles translate the user's drag input into character locomotion. Once the user touches and drags a handle, the system computes the target destination and updates the character's pose and animation state in real-time. Collision bounds on the platform restrict movement to allowable regions. If the character reaches the target or the user stops dragging, movement ceases, and the interaction state resets.

\textbf{Attach a Prop.}
The system also supports prop attachment using virtual ``hand zones'' attached to each character’s skeleton (Figure~\ref{fig:interface}c). Objects brought near a hand zone within a proximity threshold will automatically attach to the corresponding hand, offering a natural interaction for storytelling scenarios such as giving or dropping items. 

 These interaction techniques combine to support an intuitive and expressive authoring experience, allowing users to choose the modality that best matches their creative intent at a given moment of authoring (DG2).

\subsubsection{Spontaneous Dialogue}

The dialogue system is designed to create the feeling of a live back-and-forth exchange  that invites exploration and improvisation (DG1). At any point, users can take control of a character by picking it up and speaking through it, while the remaining characters are played by the AI. When controlled by the AI, each character remains aware of who it is speaking to and responds directly, allowing conversations to unfold fluidly.  If users are not satisfied with an AI character's response, they can simply grab the character and override its previous line by voicing their desired response. This encourages trying new ideas without the fear of failure (DG1). 

\textbf{User Speech.} Users input speech by grabbing and holding a character. This design enables users to directly voice the character they are holding, offering a natural and intuitive method for taking over control of the character. Recognized speech is logged and shown in the dialogue history in the scene, tagged with the timestamp and the character name from which the user is speaking from.

\textbf{AI Speech.} Character responses are generated dynamically by an LLM and grounded in the narrative context. When AI speech is activated, the character will display a speech bubble and orient its body towards the user-controlled character (Figure~\ref{fig:interface}d). This response is shaped by both the content of the user's utterance and narrative metadata such as character role, emotional state, and scene context. The AI speech operates in two modes:

\begin{itemize}
    \item \textit{Reactive Speech:} When a user finishes voicing a character, the system triggers a context-aware response from the corresponding AI character by first checking if their line includes another character’s name, which allows explicit targeting. If no name is mentioned, the system uses spatial orientation: whichever character the speaker is facing to infer the addressee.  
    
    \item \textit{Proactive Speech:} In moments of user inactivity, the AI takes initiative to progress the story. If no input (speech, grab, or movement) is detected for 10 seconds, the system interprets this as inactivity and allows a character to speak autonomously.  This proactive mode enables characters to prompt the user, suggest narrative directions, and keep users engaged and curious about what will happen next. This maintains narrative momentum to cultivate a sense of open-ended play (DG1).
\end{itemize}

Both reactive and proactive responses are orchestrated by a centralized character speech handler that interfaces with a multi-agent pipeline. Characters leverage predefined personality tags and dynamic scene cues to maintain narrative coherence across exchanges. 

\subsubsection{Story Marble \& Assembly}

 The system captures the user's interactions and translate them into story events (DG3). We use \textit{Story Marble} as the visual representation of meaningful story events extracted from spatial interactions and dialogue. 
 Marbles are designed to be small, self-contained units that can be picked up, grouped, or discarded to foster playfulness, emphasizing that story events are interactive \textit{tangible bits} \cite{ishii1997tangible} rather than compiled code blocks (DG1). 
Users can pick up marbles and arrange them in a story timeline to re-sequence the story as an alternative to their chronological play (Figure~\ref{fig:interface}e). Upon picking up a marble, the system replays the 3D scene and dialogue with characters on their marks when the event is captured.  Interactions with marbles are designed to encourage users to try alternatives. The metaphor also aligns with the embodied improvisational spirit of the system (DG1): users tinker with spatial proxies instead of assembling a script directly.

Each marble contains a card that shows a description and a snapshot of the event.  This card is a summarization of \textit{Intent Frame}, a modular story unit that captures rich story data such as narrative intent, story tension, characters, and user actions (DG3). During a session, user actions such as grabbing, talking, or moving generate corresponding \textit{Intent Frames}. These are spawned in real-time by a centralized manager that listens for interaction events, tags them with appropriate context, and queues them in a session-level log. This log not only supports replay and analysis but also serves as the foundation for subsequent narrative structuring.

 Each Intent Frame encodes a semantically meaningful abstraction of the user's action, including its inferred intention, emotional tone, and story context. For example, picking up a character and saying \textit{``Have you heard? The magic lamp we’ve been searching for has finally been found — but it's in the hands of our greatest rival''} might be distilled into an \texttt{Inciting Incident} with linked characters, action, and prop. This structure allows the system to fluidly support both improvisational play (DG1) and structured story building (DG3).



\subsubsection{Story Generation}
Once arranged, the user can export the story as a fully structured artifact—either a synopsis or a screenplay—capturing the assembled narrative beats in a form directly usable for writing or performance. This export includes the ordered sequence of marbles, dialogue history, and metadata about the session (e.g., duration, export time, and interaction types). 
The interface supports two types of output: a \textit{Story Summary} and a \textit{Screenplay}: 

The \textbf{Story Summary} shows a three-paragraph synopsis. The summary highlights story setup, development, and resolution, drawing on specific quotes and character intentions derived from the session data. 
This provides users with a concise, readable account of the narrative that can be used for reflection or further editing.

The \textbf{Screenplay} transforms the entire session log into a screenplay-style output, complete with scene headers, speaker attributions, and formatted dialogue. The tool uses session metadata, such as the specified environment and the relative activity levels of different characters, to emphasize setting and identify central characters.  Dialogue history is serialized into structured text with formatting suitable for iterative authoring or print export.



\subsection{System Pipeline}

\begin{figure*}[h]
  \centering
  \includegraphics[width=0.98\linewidth]{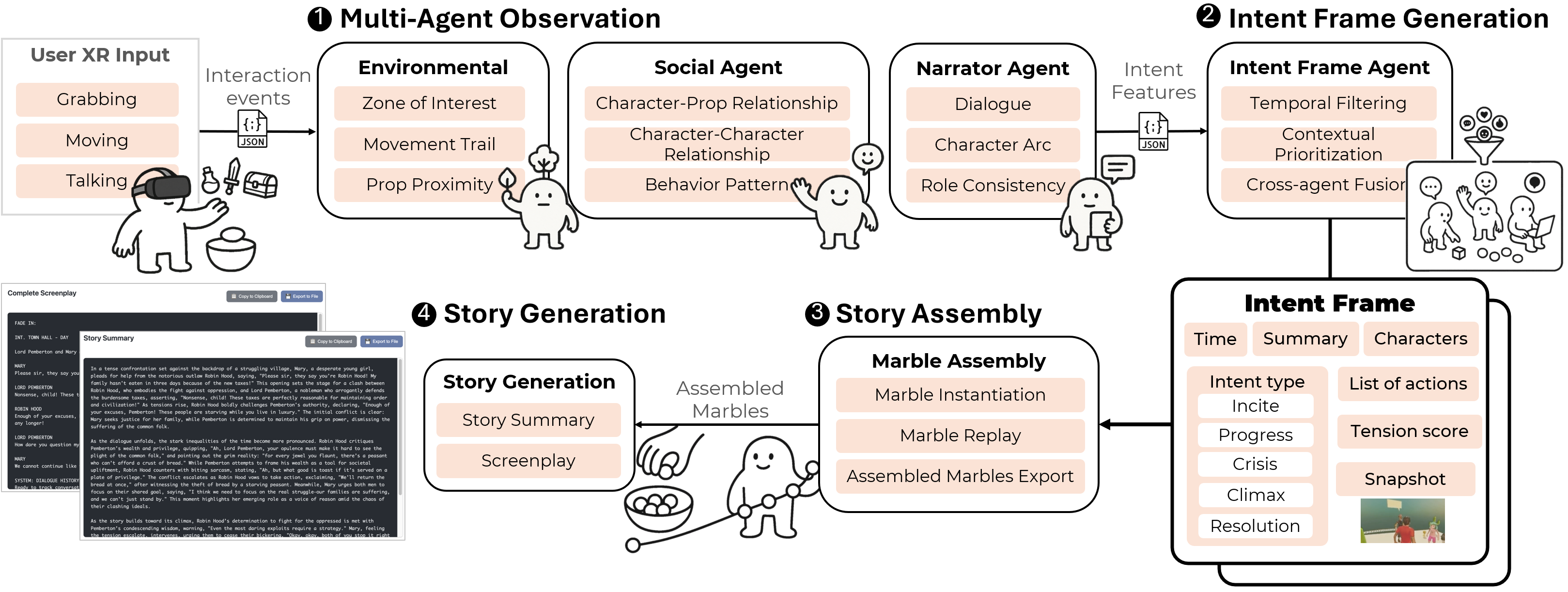}
  \Description{Flowchart-style diagram showing PlayWrite’s pipeline. On the left are user inputs (grabbing, moving, talking). In the center are three types of agents (Environmental, Social, Narrator). Arrows lead to Intent Frame generation and then to Story Assembly and Story Generation. Small icons of characters and objects appear alongside each stage.}
  \caption{System pipeline of \textit{PlayWrite}. 
(1) Multi-agent observation collects user inputs (speech, grabbing, moving) and interprets them through Environmental, Social, and Narrator agents. 
(2) The Intent Frame Agent fuses these into higher-level features through temporal filtering, contextual prioritization, and cross-agent fusion. 
(3) Intent frames are assembled into marbles, which can be replayed, reordered, or exported. 
(4) Story generation produces a coherent story summary and screenplay from the assembled marbles.}
  \label{fig:agent-context-example}
\end{figure*}

To enable dynamic, context-aware storytelling, our system employs a multi-agent pipeline that observes the scene, abstracts interaction semantics, and converts them into \textit{Intent Frames}. At the core of this pipeline is a modular multi-agent structure that continuously monitors various dimensions of the environment and feeds its observations into an Intent Frame Agent. Rather than relying on low-level input events, the system works with these semantically rich units, which serve as the basis for marble assembly and story generation.

\subsubsection*{1. Multi-Agent Observation}

To robustly interpret the evolving narrative landscape, our system deploys a modular architecture composed of three dedicated agents—\textit{Environment Agent}, \textit{Social Agent}, and \textit{Narrator Agent}—each continuously observing a distinct contextual layer of the 3D scene. These agents operate in parallel and asynchronously report semantically meaningful features to the central \textit{Intent Frame Agent}, which fuses them into structured \textit{Intent Frames}.

\paragraph{Environment Agent}
The \textit{Environment Agent} monitors the physical configuration and spatial dynamics of the scene. It implements continuous polling routines that track:

\begin{itemize}
    \item \textbf{Zones of Interest }: Monitors the location of the characters and props, whether they are placed in certain zones with tags (e.g, Hiding Zone) to communicate the social dynamic with the Intent Frame Agent (Figure~\ref{fig:agent-context-example}a).
    
    \item \textbf{Movement Trails}: Logs recent position changes to infer behavioral patterns such as ``character moved from position A to B,'' or ``object thrown from position A to B.''
        
    \item \textbf{Prop Proximity}: Computes pairwise distances between props and characters, enabling detection of interactions like ``character approaching the campfire'', or ``character guarding the treasure''. Identifies proximity-based groupings of interactable items, such as ``gathering loot, gold, and treasure''. 
\end{itemize}

These features are packaged as events and asynchronously buffered. The \textit{Intent Frame Agent} selectively consumes these observations based on their Environmental Salience $S_e$, which is a heuristic score that quantifies the narrative importance of a physical event, allowing the system to distinguish between deliberate authorial intent and incidental spatial changes. This score is determined by several factors implemented in the \textit{Environment Agent}. For instance, an event's salience increases with significant movement, where a character or object surpasses a predefined distance threshold (\texttt{m\_MovementThreshold}). Similarly, proximity events are highly salient, such as when a character moves within an interaction radius of a key prop or another character. Finally, salience is also determined by zone-based meaning, where entering a predefined \textit{Zone of Interest} is an inherently meaningful event that provides immediate narrative context.

\paragraph{Social Agent}
The \textit{Social Agent} captures behavioral and inter-character dynamics by analyzing:

\begin{itemize}
   \item \textbf{Character-Prop Relationships}: Detects interactions between characters and nearby props, characters approaching one another, or direct manipulation. This helps surface moments such as a character reaching for a specific object or remaining near a symbolically relevant item.
    
    \item \textbf{Character-Character Interactions}: Identifies behavioral correlations between characters or between users and characters, such as coordinated movement and mutual orientation.
    
    \item \textbf{Behavioral Patterns}: Extracts recurrent sequences of actions and emotions that characters exhibit in response to specific situations (e.g, being threatened by a gun) - revealing interaction intent from direct manipulation.
    
\end{itemize}

Each social context is encoded as events, enriched with role-based implications (e.g., ``supportive partner,'' ``dominant antagonist'') for narrative alignment. The system prioritizes these events based on their Social Novelty ($S_s$), a heuristic score that quantifies the newness of a social event to distinguish meaningful developments from incidental interactions. As implemented in the \textit{SocialAgent}, novelty is highest for first-time interactions, such as when two characters come within proximity of each other for the first time. To filter out repetitive noise, the novelty of a subsequent interaction between the same characters is significantly reduced if it occurs within a short time window (a 5-second cool-down), ensuring that only sustained or newly re-initiated encounters are flagged as narratively important.

\paragraph{Narrator Agent}
The \textit{Narrator Agent} acts as the memory backbone of the narrative engine, managing temporal coherence across the session. It maintains:

\begin{itemize}
    \item \textbf{Dialogue}: Maintains complete conversation history, enabling callbacks and re-visitations of prior topics.
    \item \textbf{Character Arcs}: Tracks emotional states, evolving goals, and unresolved tensions for each character.
    \item \textbf{Role Consistency}: Modulates AI responses according to internal narrative variables such as character motivation, alliance, and moral alignment.
\end{itemize}

This agent ensures that all emergent actions align with both short-term plausibility and long-term continuity, preserving thematic integrity across scenes. To maintain this long-term continuity, the \textit{Narrator Agent} uses the narrative parameters defined in the \texttt{StoryRoleConfiguration} as its guiding framework. This configuration file serves as a blueprint for the story, outlining the thematic guidelines such as the location, time, and core personality traits for each character. The \textit{Narrator Agent} consistently references this blueprint when generating narrative text. By doing so, it ensures that the dialogue and character arcs remain thematically coherent and aligned with the configuration, regardless of whether the scene is structured as goal-driven or open-ended.

\begin{figure*}[h]
  \centering
  \includegraphics[width=0.6\linewidth]{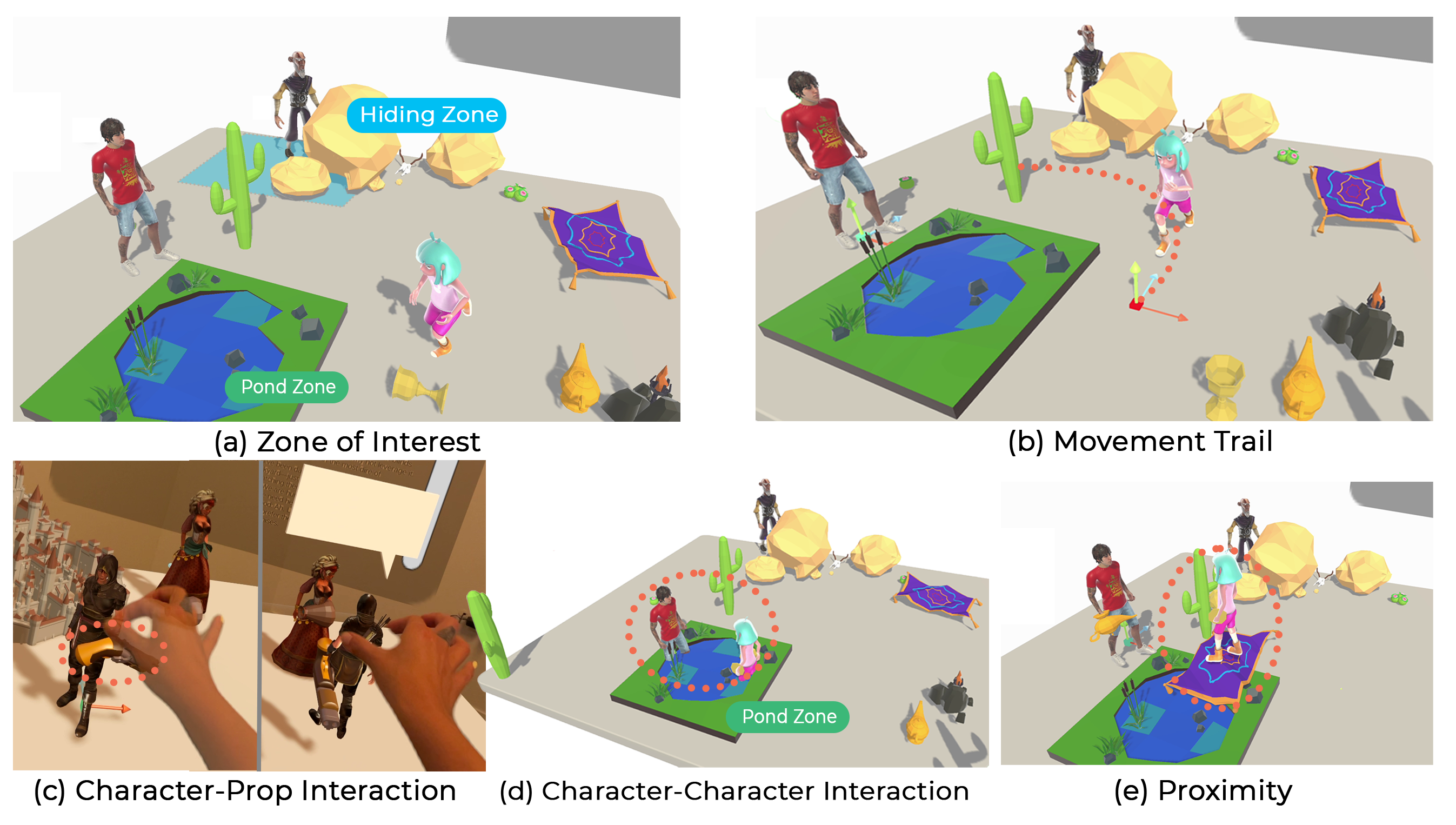}
  \Description{Exemplary interactions: sequence of frames showing a user arranging characters, triggering speech, assembling Story Marbles, and exporting a screenplay.}
  \caption{Examples of semantically meaningful features captured during play and encoded into Intent Frames. 
(a) \textbf{Zone of Interest:} a character placed within a predefined hiding zone, and therefore, his action and speech will not be noticed by other characters. 
(b) \textbf{Movement Trail:} the movement of a character is tracked to capture directionality and pacing. 
(c) \textbf{Character–Prop Interaction:} a gun is attached to signal a threat. Other characters will react to seeing the gun in hand.  
(d) \textbf{Character–Character Interaction:} relative positioning and engagement between characters are recorded as two characters relaxing together in the pool. 
(e) \textbf{Proximity:} spatial closeness between the character and a magic carpet indicates plot progression. 
Together, these features provide structured cues that the system abstracts into Intent Frames for narrative interpretation.}
  \label{fig:agent-context-example}
\end{figure*}

\subsubsection*{2. Intent Frame Generation}


All three contextual agents—the Environment Agent, Social Agent, and Narrator Agent—operate asynchronously in real time, continuously observing the interactive scene. Each agent emits semantically tagged observations in the form of \textit{IntentFeatures}, which include metadata such as \texttt{actor}, \texttt{target}, \texttt{location}, \texttt{timestamp},  \texttt{semanticLabel}, and \texttt{confidence}. 

In this context, the \texttt{actor} is the primary entity initiating the event (e.g., the character being moved), while the \texttt{target} is the entity being acted upon (e.g., another character or a prop). The \texttt{location} and \texttt{timestamp} record the precise spatial and temporal context of the observation. The \texttt{semanticLabel} provides a layer of interpretation; it is a descriptive tag (e.g., character\_speech, prop\_proximity) that translates a raw system event, such as a change in a Vector3 coordinate, into a narratively meaningful concept. Finally, \texttt{confidence} is a score representing the agent's certainty about its interpretation, allowing the system to weigh direct, unambiguous user inputs more heavily than inferred observations.
These features are then combined downstream into higher-level \textit{Intent Frames}, which represent semantically meaningful narrative actions.  This structured pipeline ensures that observations from different agents remain interoperable, time-aligned, and semantically rich for narrative interpretation.

The \textit{Intent Frame Agent} functions as a central fusion module that performs \textbf{temporal filtering}, \textbf{contextual prioritization}, and \textbf{cross-agent fusion}. First, it applies temporal filtering to suppress trivial or redundant input events (e.g., jittery object movements or repeated low-salience cues). Then, it dynamically prioritizes observations based on their relevance to active narrative arcs, spatial transitions, or user-driven interaction patterns. Finally, it merges co-occurring observations across agents into unified intent frames—for example, if an Environment Agent reports movement and the Social Agent detects a shift in inter-character relation, this might be fused into a single high-level description like ``approach with anger.''

Each \textit{Intent Frame} is ranked using a hybrid scoring function:
\[
R = w_e E + w_s S + w_n N
\]
where $E$ is \textit{Environmental Salience} (e.g., object displacement $> 0.5$m), $S$ is \textit{Social Novelty} (e.g., first-time character interaction), and $N$ is \textit{Narrative Progression Likelihood} derived from the Narrator Agent’s logs. The weights $w_e$, $w_s$, and $w_n$ are adjusted dynamically at runtime based on recent event history: for instance, down-weighting repeated environmental motions or up-weighting novel social interactions—to balance variety and prevent repetitive patterns. Events are stored in a priority queue, and once a configurable threshold (e.g., $N = 5$) is reached, or after a timeout window ($T = 1s$), the \textit{Intent Frame Agent} initiates a commit.

Committed intent frames are instantiated as C\# objects and include a ranked list of merged actions, characters, time, tension score, and a resolved \texttt{intentType} (e.g., ``negotiation escalation''). The \texttt{tension score}, which serves as a direct measure of conflict, is a numerical rating from 1 (calm) to 10 (extreme conflict) representing the scene's dramatic intensity. It is computed by an LLM that analyzes the actions and dialogue for emotional stakes and narrative conflict. The \texttt{intentType}, in turn, is a classification that categorizes the scene according to its narrative function within a traditional three-act structure (e.g., \textit{IncitingIncident}, \textit{RisingAction}, \textit{Climax}). This classification is grounded in established narrative theories articulated by authors such as Robert McKee \cite{mckee1997substance}. This modular architecture allows multiple agents to act independently while contributing to a semantically unified representation of what matters in the scene. The generated \textit{Intent Frames} are stored in the background while the user continues to play.

\subsubsection*{3. Story Assembly}  
Once the play ends, the user enters the Story Assembly stage, where the \textit{Intent Frames} that were generated are visually instantiated as \textit{Story Marbles} on a timeline. Each marble serves as a manipulable proxy for a narrative moment, containing key data from its Intent Frame, such as a summary, the characters involved, and a visual snapshot of the scene. Users can rearrange the marbles on the timeline, directly altering the sequence of story beats. The story assembler treats the assembled marbles as the canonical sequence and passes them to the LLM-based generator in that order. When adjustments are made (e.g., moving a confrontation slightly earlier or deleting a redundant beat), the resulting summaries remain locally coherent, with the model inferring reasonable transitions between beats. 


\subsubsection*{4. Story Generation}  
In this final stage, the system takes the user's arranged sequence of assembled marbles and transforms it into a narrative. The ordering of the marbles affects the final output, ultimately shaping how the story evolves.

The system supports multiple output formats, including a high-level Story Summary that provides a concise synopsis and a formatted Screenplay with scene headers and dialogue. This design enables a high-fidelity record of the orchestrated narrative, maintaining coherence across spatial, social, and storytelling dimensions.

\subsection{Technical Implementation}

Our system is built in Unity 6 and deployed on Apple Vision Pro using the PolySpatial rendering backend to support mixed-reality interfaces. We integrate a real-time character interaction system with LLM-backed response generation, maintaining latency bounds under 2 seconds for dialogue turns. Each character is initialized with a configurable token budget (typically 512--2048 tokens) and a unique system prompt containing personality, goals, and narrative context. Prompts are dynamically assembled at runtime using a truncated interaction history, filtered object and character context, and global narrative summaries. Token estimation is handled client-side with a lightweight tokenizer to avoid exceeding model limits. We use GPT-4-turbo for character dialogue and intent interpretation. Responses are streamed and interpolated for pacing and emotional prosody, then rendered as spatial audio anchored to the speaking character. All outputs are logged as timestamped marbles in JSON format, with each marble preserving entity associations and semantic role tags for downstream narrative analysis and replay.

\begin{figure*}[t]
  \centering
  \includegraphics[width=\linewidth]{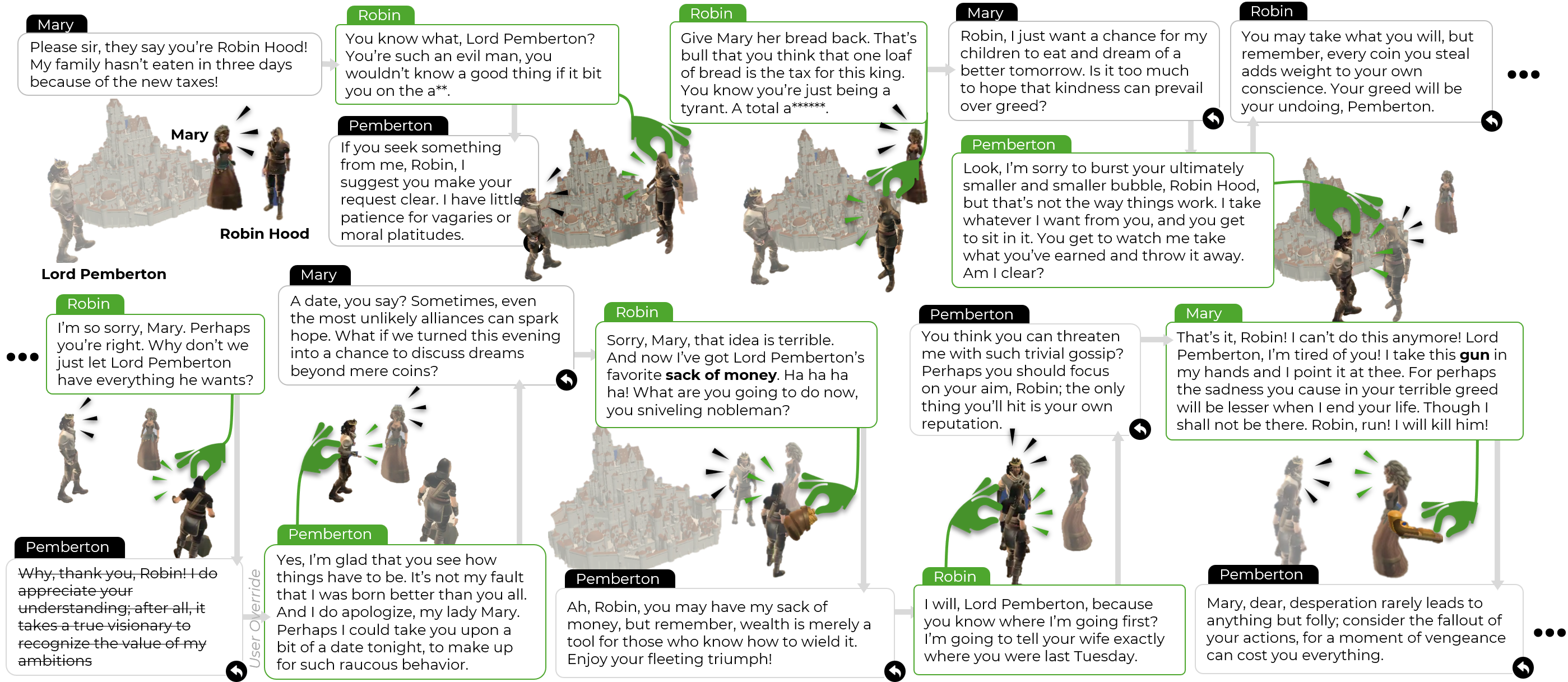}%
  \Description{Exemplary interactions: sequence of frames showing a user arranging characters, triggering speech, assembling Story Marbles, and exporting a screenplay.}
  \caption{ Recorded example workflow of Robin Hood extracted from a user's play (P6, open-ended task). Green-labeled boxes show user speeches, and black-labeled boxes show AI responses demonstrating two conversation segments: (first row) Mary pleads with Robin Hood for help. The user, controlling Robin, confronts Lord Pemberton. (second row) The story develops as the user switches characters and overrides AI responses to introduce a plot twist. Complete dialogue and synopsis can be found in Appendix E and F.}
  \label{fig:workflow}
\end{figure*}

\section{Example Workflow: Story Authoring in PlayWrite}
\label{sec:workflow}


 To illustrate how PlayWrite supports embodied story authoring, we follow Lukas, an improvisation teacher and dungeon master, through a typical session. Upon launching the system, he enters a mixed-reality stage containing three characters—\textit{Mary}, \textit{Lord Pemberton}, and \textit{Robin Hood}—alongside props such as a pistol, a sack of gold and a chalice.

The scene begins with Mary pleading for help for her starving family, and Robin confronting Lord Pemberton about exploiting the poor (Figure~\ref{fig:workflow}, first row). These early beats are driven primarily by Lukas’s embodied performance: moving characters toward each other, placing them into proximity, and voicing their lines in-role: \textit{“...You’re such an evil man, you wouldn’t know a good thing if it bit you on the a**”}. The system then responds with AI-generated dialogue that heightens the conflict. At one point, Lord Pemberton replies: \textit{“...I have little patience for vagaries or moral platitudes”}. Lukas continues the confrontation by switching between characters, just as he did when playing with dolls as a child, alternating roles by picking up different figures rather than issuing system commands. 

As the conflict escalates, the character of \textit{Mary} joins the exchange when her figure is brought into the interaction: \textit{``Robin, I just want a chance for my children to eat and dream of a better tomorrow''}. At this point, instead of picking up Robin to comfort \textit{Mary}, Lukas grabs \textit{Lord Pemberton} and moves him to the center of the scene while speaking in-role: \textit{“Look...I take whatever I want from you, and you get to sit in it. You get to watch me take what you've earned and throw it away. Am I clear?”} Curious how the system would respond, he pauses. The AI then takes over \textit{Robin Hood}: \textit{“You may take what you will, but remember, every coin you steal adds weight to your own conscience. Your greed will be your undoing, Pemberton.”}

As the interaction goes on, authorship alternates between Lukas and AI continuation (Figure~\ref{fig:workflow}, second row). When Lukas is unsatisfied with AI responses, he simply overrides the dialogue by grabbing and voicing out the AI-controlled character. Lukas plays Robin, stealing a sack of gold: \textit{``And now I've got Lord Pemberton's favorite sack of money. Ha ha ha ha! What are you going to do now, you sniveling nobleman?''} The AI responds with \textit{Lord Pemberton}: \textit{``Ah, Robin, you may have my sack of money, but remember, wealth is merely a tool for those who know how to wield it. Enjoy your fleeting triumph''}. 
Seeing \textit{Robin}’s threat not working on \textit{Lord Pemberton}, Lukas adjusted his way of playing by mocking  \textit{Pemberton}:\textit{“I will, Lord Pemberton, because you know where I'm going first? I’m going to tell your wife exactly where you were last Tuesday”}. This twist about \textit{Pemberton}’s private life is entirely user-created, and the AI subsequently folds it into \textit{Pemberton}’s dismissive, self-justifying replies: \textit{``You think you can threaten me with such trivial gossip?...The only thing you’ll hit is your own reputation.''}.

The escalation around the gun is likewise a blend of physical and verbal authorship. Lukas grabs the pistol and gives it to \textit{Mary} and voices her breaking point: \textit{“That’s it, Robin! I can’t do this anymore... I take this gun in my hands and I point it at thee...”}, while the AI contributes \textit{Lord Pemberton}’s calm, condescending reaction that frames her act as “desperation” and “folly.” The scene resolves with Mary’s final refusal to kill him, again spoken by Lukas, followed by an AI response that keeps  \textit{Pemberton} oblivious and smug.


Once satisfied with the performance, Lukas brings the story marbles into view. Lukas examines marbles displaying concise summaries; lifting one replays the moment in space, restoring character poses and highlighting the corresponding dialogue.
From this play, Lukas gets the following marbles: 

\begin{itemize}
    \item \textbf{Marble 1 – Mary pleads with Robin Hood} \\
    \textit{Trigger:} Mary moves toward Robin and speaks. \\
    \textit{Authorship:} character movement + user speech.
    \item \textbf{Marble 2 – Mary implores Robin to save her family} \\
    \textit{Trigger:} conversation between Mary and Robin. \\
    \textit{Authorship:} user speech + AI response.
    \item \textbf{Marble 3 – Robin confronts Lord Pemberton in Sherwood} \\
    \textit{Trigger:} Robin moves and addresses Pemberton. \\
    \textit{Authorship:} user speech + AI response.
    \item \textbf{Marble 4 – Robin steals Pemberton’s gold bag} \\
    \textit{Trigger:} Robin grabs the gold bag prop and confronts Pemberton. \\
    \textit{Authorship:} user manipulation + user speech + AI response.
\end{itemize}

Lukas rearranges the marbles to restructure the narrative—placing the confrontation earlier and grouping rising-action beats. When the story feels right, he selects \textbf{Export the Play}.
PlayWrite exports the assembled marble sequence as a structured JSON log. Using a web interface, Lukas generates a synopsis (Appendix E) and a formatted screenplay (Appendix F). He reviews the generated text, appreciating that a subtle vocal tone was preserved in the export before copying the draft into his own writing environment for further refinement.



\section{Exploratory User Study}

We conducted an exploratory user study with 13 participants to evaluate \textit{PlayWrite}. The goals of the study were to:
(1) understand how writers use and perceive the play-based, multimodal interface for story creation in XR,
(2) evaluate the effectiveness of the multi-agent pipeline in translating embodied play into narratively meaningful Intent Frames that reflect authorial intent, and
(3) explore the benefits and drawbacks of a co-creative storytelling workflow where narrative emerges from a playful, embodied partnership between the author and the AI.

\subsection{Participants}

We recruited 13 participants  (8 female, 4 male, 1 non-binary) aged between 21 and 45 (\textit{M} = 31.2, \textit{SD} = 6.8), with diverse backgrounds in writing and narrative creation  (Table~\ref{tab:participant_backgrounds}). Participants included published authors, fiction writers, hobbyists, and game designers. Educational backgrounds ranged from Bachelor’s (6), Master’s (5), to PhDs (2), and most reported regular writing habits—ranging from weekly to several times a week.

\begin{table}[h]
\centering
\small
\caption{ Participant backgrounds, mapped to the anonymized IDs (P1--P13) used when reporting quotes and findings.}
\Description{Table listing each participant with demographics such as writing background and primary storytelling medium.}
\setlength{\tabcolsep}{3pt}
\renewcommand{\arraystretch}{1.1}
\begin{tabular}{l p{0.55\columnwidth} p{0.30\columnwidth}}
\toprule
\textbf{ID} & \textbf{Writing Background} & \textbf{Primary Writing Format} \\
\midrule
P1  & Songwriting, interactive media storytelling, fantasy, science fiction, screenwriting, game narrative, poetry & Scripts/screenplays  \\
P2  & Poetry, journals, literary fiction & Short stories  \\
P3  & Fiction writing, poetry, literary fiction & Short stories  \\
P4  & Prose fiction, fantasy, science fiction, poetry & Short stories  \\
P5  & Science fiction writing, screenwriting, game narrative & Short stories \\
P6  & Dungeon master, improv comedy, game narrative, interactive fiction & Improv comedy and D\&D narratives \\
P7  & Recreational creative writing, D\&D game mastering, fantasy, romance, comedy & Short stories \\
P8  & Graphic novel writing & Short stories  \\
P9  & Songwriting, fantasy, poetry & Songs  \\
P10 & Songwriting, branching narratives, poetry & Short stories and branching narratives \\
P11 & Creative writing, academic writing, fantasy & Short stories  \\
P12 & Worldbuilding, game narrative, interactive fiction, fantasy, science fiction & Worldbuilding stories and documents  \\
P13 & Essay and article writing about life, humanity, and technology & Short stories \\
\bottomrule
\end{tabular}
\label{tab:participant_backgrounds}
\end{table}

Genre preferences spanned science fiction, fantasy, poetry, graphic novels, and game narratives, while primary writing formats included short stories, branching narratives, and world-building documents. The majority identified their writing style as “hybrid” (i.e., mixing pre-planned outlines with improvisational flow), though a few leaned toward purely discovery-driven writing.
Participants also varied in their technological fluency: 5 considered themselves early adopters of creative tech, 4 rated themselves as eager but cautious, and 4 expressed limited comfort. 

\subsection{Procedure}

\begin{figure*}[t]
  \centering
  \includegraphics[width=0.98\linewidth]{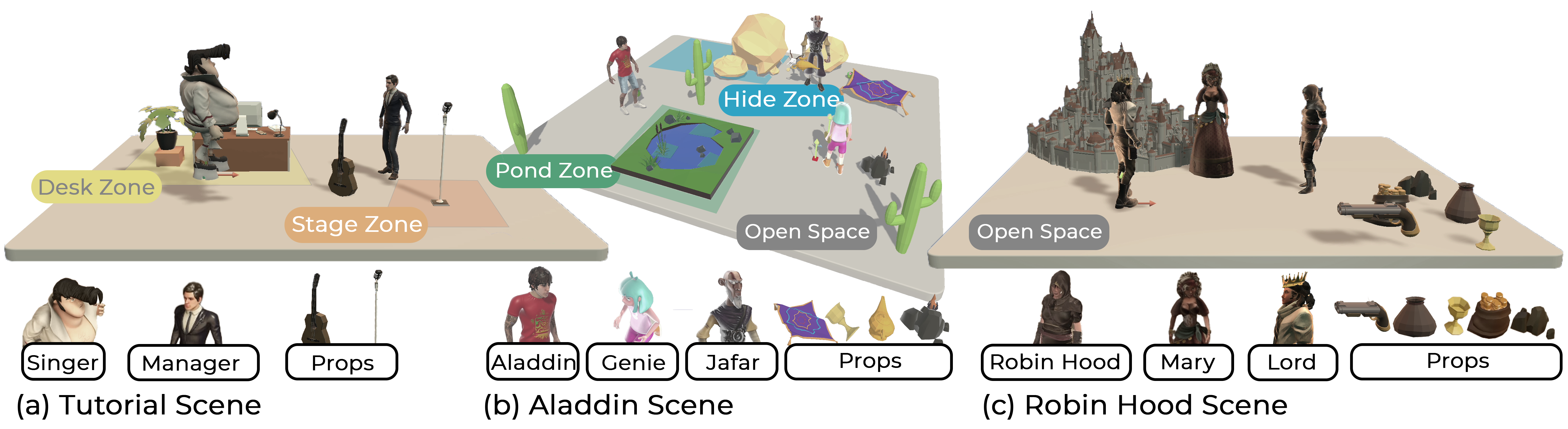}
  \Description{Three different scenes shown side-by-side. Each has unique characters, props, and a distinct environment theme. The images represent the pre-authored starting scenes used in the study.}
  \caption{Three distinct scenes used in the user study. 
    (a) \textbf{Tutorial Scene:} A simple environment depicting an interview between two characters with a desk zone and stage zone, including an Elvis impersonator, a hiring manager, and basic props. This scene was used to introduce participants to the system’s core interactions (grabbing, moving, speaking, attaching props).  
    (b) \textbf{Aladdin Scene:} A desert-themed environment with pond and hide zones, featuring Aladdin, Genie, and Jafar alongside interactive props. Participants were encouraged to improvise and stage interactions using familiar characters. They were presented with an end goal to reach.
    (c) \textbf{Robin Hood Scene:} A castle environment with an open space zone, featuring Robin Hood, Mary, and Lord Pemberton, along with thematic props. Participants were instructed to use the characters and the props for an open-ended narrative exploration.
    Together, these three scenes provided varying levels of narrative structure, scaffolding participants from practice to open-ended improvisation.}
  \label{fig:study-scene}
\end{figure*}

Each study session lasted approximately 90 minutes and was conducted in a mixed reality environment using an Apple Vision Pro headset. Participants were guided through a series of narrative tasks. This study was reviewed by and received approval through our institution’s internal ethics review process.

\subsubsection{Tutorial and On-boarding (10 minutes)}
Participants were first introduced to the system with a tutorial to build fluency with the system's core interactions. This scene featured two characters—an Elvis impersonator and a job interviewer—in an office environment with a desk and stage zone (Fig~\ref{fig:study-scene}a). Props like a guitar and a microphone were provided to encourage playful experimentation with grabbing, moving, and invoking AI responses.

\subsubsection{Closed-Ended Narrative Authoring (10 minutes)}
 For the first task, participants were given a defined beginning and a required ending within this scene, which was based on the story of \textit{Aladdin} (Fig~\ref{fig:study-scene}b). The scene included the characters Aladdin, Jafar, and the Genie, along with props like a magic lamp and carpet in a playground with hide areas and open space. Participants were instructed to guide the story from a celebratory start to a fixed outcome where Jafar successfully seizes the lamp, leaving Aladdin betrayed and vowing revenge.

\subsubsection{Open-Ended Narrative Authoring (10 minutes)}
The second task focused on open-ended, improvisational storytelling to assess support for divergent creativity. This scene featured Robin Hood, Mary, and Lord Pemberton in an open space with thematic props like a sack of gold and a pistol (Fig~\ref{fig:study-scene}c). Although the story began with a specific inciting incident (Mary's bread being confiscated), participants were not given a target outcome. Instead, they were encouraged to freely explore character motivations and plot twists to craft a story in any direction they found compelling.

\label{marbleassembly}
\subsubsection{Marble Assembly and Story Review (10 minutes)}

At the end of each task, participants reviewed the Story Marbles that the system generated from their play. They examined and reordered the Story Marbles in the timeline. Once they  were satisfied with the marble order, they notified the experimenter, who exported the JSON file which logged the session. This JSON was then imported into a custom web interface which generated a short synopsis and a screenplay version of their play.  The participants reviewed the generated story and  completed a post-task survey. 

\subsubsection{Post-Study Questionnaire and Interview}

At the end of the study, participants completed the Creative Support Index (CSI) \cite{10.1145/2617588} questionnaire and a post-study questionnaire that focused on the overall user experience.
This was followed by a semi-structured interview focusing on the narrative control, the AI's role, system usability, and the interplay between physical interaction and story co-creation. These reflections provided insight into the experiential and functional attributes of the system. The full list of semi-structured interview questions can be found in the Appendix.

\subsection{Data Analysis}

We collected post-task, post-study questionnaires, audio and video screen recordings from the scene, and the corresponding text-to-speech transcriptions. After each session, the two researchers who participated discussed their observations and exchanged notes. After all studies had been completed, the first author conducted thematic clustering \cite{blandford2016qualitative}. The clusters were then further elaborated and refined by the co-authors in conjunction.

\subsection{Results}


\begin{table}[h]
\centering
\caption{Creativity Support Index (CSI) Results (N=13)}
\begin{tabular}{lcc}
\toprule
Factor & Mean & SD \\
\midrule
Expressiveness       & 6.35 & 0.77 \\
Enjoyment            & 6.15 & 0.80 \\
Immersion            & 5.85 & 1.41 \\
Results Worth Effort & 5.85 & 1.07 \\
Exploration          & 5.38 & 1.43 \\
Control              & 3.92 & 1.93 \\
\bottomrule
\end{tabular}
\label{tab:csi}
\Description{Table showing Creative Support Index (CSI) subscale scores (e.g., Enjoyment, Exploration, Control).}
\end{table}

\subsubsection{Questionnaire Results}

Overall, participants rated \textit{PlayWrite} highly across CSI dimensions, with \textbf{Expressiveness} (M=6.35) and \textbf{Enjoyment} (M=6.15) emerging as the strongest factors. Participants also reported positive experiences with \textit{Immersion} (M=5.85), \textit{Exploration} (M=5.38), and \textit{Results Worth Effort} (M=5.85). By contrast, perceptions of \textit{Control} were more mixed (M=3.92), suggesting that while users felt highly expressive and engaged, they did not always perceive fine-grained narrative control as central to the experience. 

\begin{figure*}[t]
    \centering
    \includegraphics[width=0.98\textwidth]{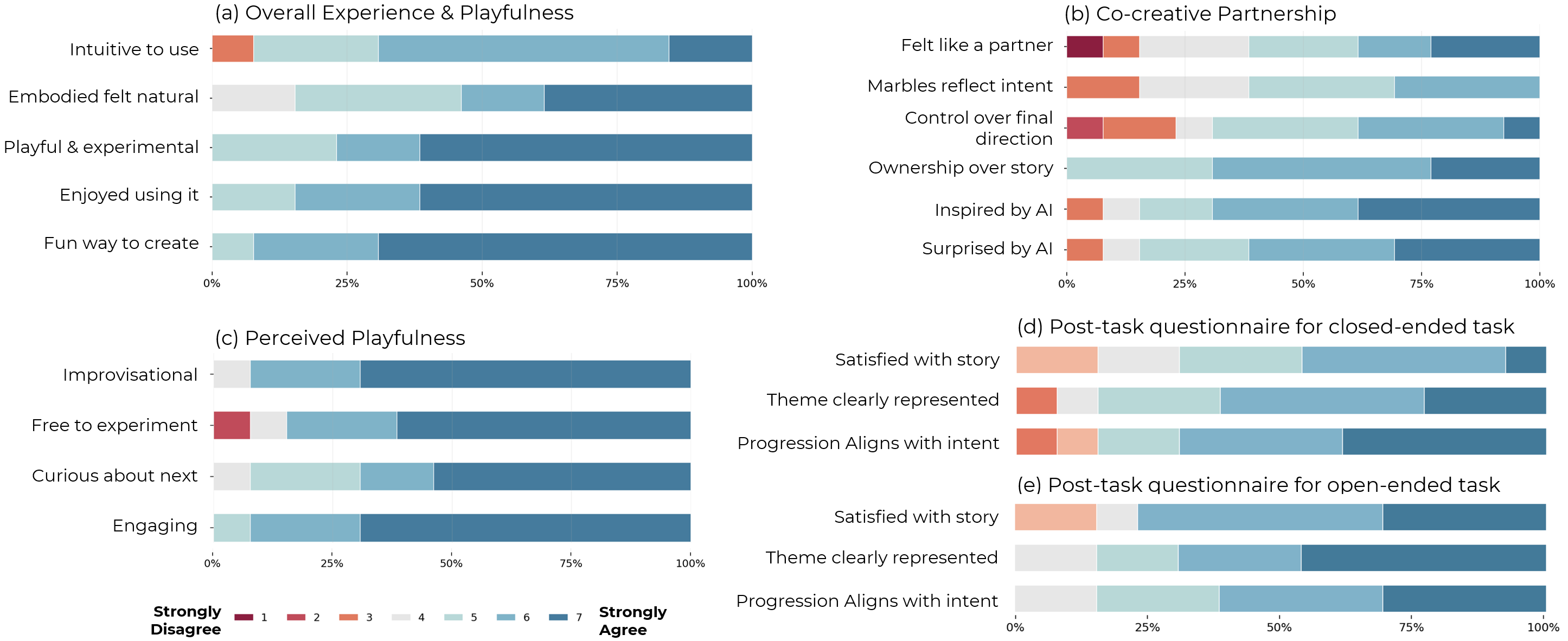}
    \Description{a) Bar chart showing participants’ responses to co-creativity and playfulness questions. The overall results are positive, with lower agreement in items related to precise control of the AI. b)Comparison bar chart across two storytelling modes: goal-oriented vs. open-ended. Each mode has multiple response bars showing different aspects of user experience; differences are visible in perceived exploration and challenge.}
    \caption{ Results of post-study questionnaires for: 
    (\textbf{a}) Overall Experience,
    (\textbf{b}) Co-creative Partnership, and 
    (\textbf{c}) Perceived Playfulness. Results of post-task questionnaires for: 
    (\textbf{d}) \textit{Closed-Ended} Task and
  (\textbf{e}) \textit{Open-Ended} Task  }
    \label{fig:questionnaire}
\end{figure*}

Beyond the CSI, post-task and post-study questionnaires provided further quantitative insight into the user experience (\ref{fig:questionnaire}). The results show a strong positive reception to the system's playful nature; participants strongly agreed that PlayWrite was a \textbf{fun way of creating stories} (M=6.54) and that it \textbf{felt playful and encouraged experimentation} (M=6.46). A majority also found the \textbf{physical, embodied interaction to be a natural way to create a story} (M=6.08).

In contrast, and consistent with the lower CSI score for \textit{Control}, participant ratings were more varied regarding narrative agency. Questions about having \textbf{sufficient control over the final narrative direction} (M=4.69) and whether the \textbf{AI’s interpretations accurately reflected their creative intentions} (M=5.00) received lower average scores, suggesting a trade-off between the playful collaboration and fine-grained authorial control.

\subsubsection{Action-Level Interaction Analysis}

\begin{figure*}[t]
  \centering
  \includegraphics[width=\textwidth]{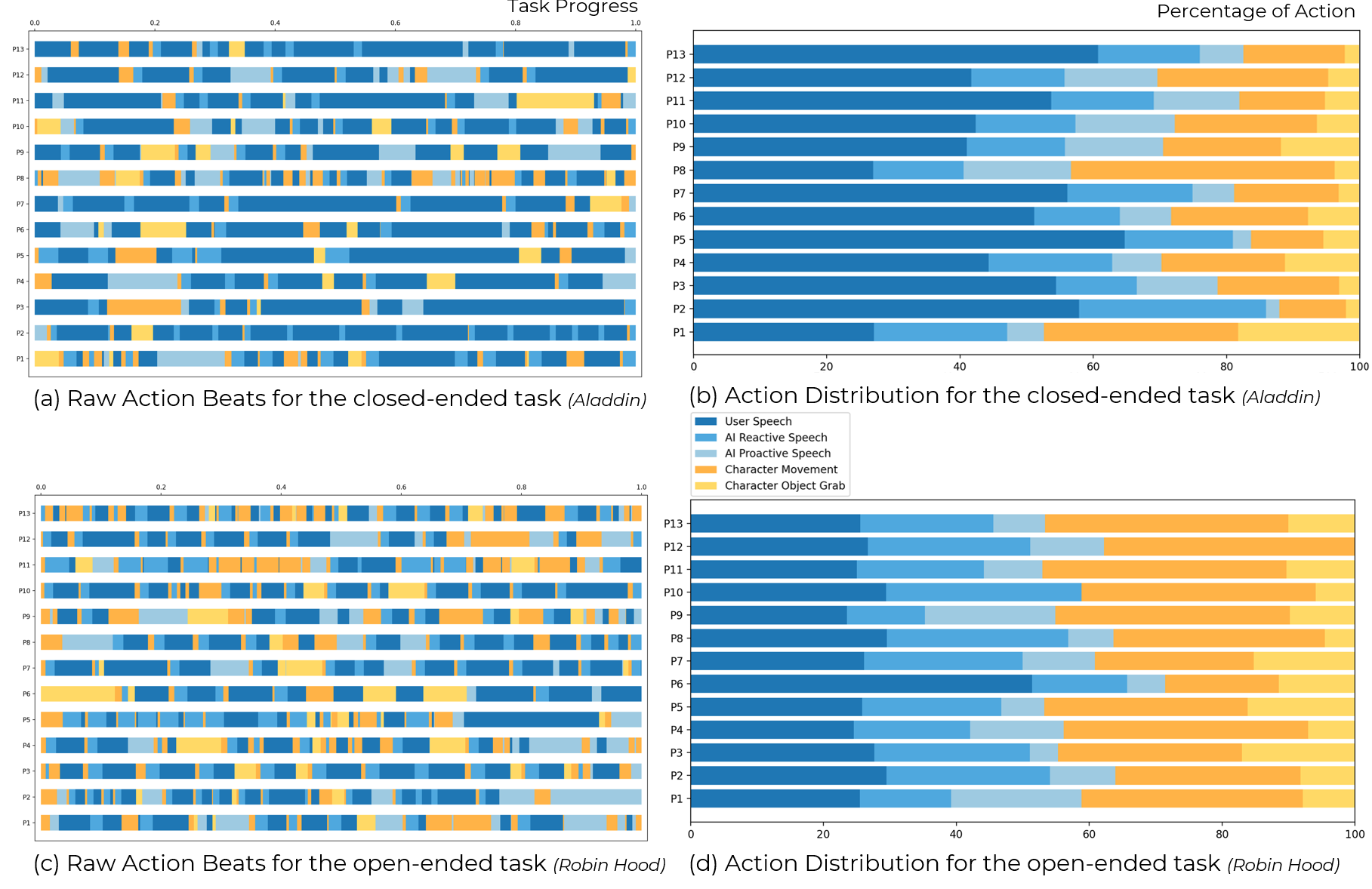}
  \Description{Stacked bar chart showing interaction types used by participants (e.g., grabbing, speech, moving). Colors differentiate interaction categories, and bars vary in length, showing differences in behavior and preference across participants.}
  \caption{Action-level interaction analysis across participants. 
  (\textbf{a}) Raw action beats for the \textit{Aladdin} scenario, showing participant timelines with color-coded action events.  
  (\textbf{b}) Stacked distribution of action types across participants in the \textit{Aladdin} scenario.  
  (\textbf{c}) Raw action beats for the \textit{Robin Hood} scenario.  
  (\textbf{d}) Stacked distribution of action types across participants in the \textit{Robin Hood} scenario.  
  Together, these visualizations highlight how participants combined speech and physical actions over time, and how relative reliance on each interaction type varied across individuals. The periods of user inactivity have been removed from the raw action beat visualizations (a, c) to focus on active interaction sequences.  
  }
  \label{fig:aladdin_robinhood}
\end{figure*}

We analyzed the distribution of interaction types across participants and sessions. Each user session was segmented into discrete action events (\textit{AI Proactive Speech}, \textit{AI Reactive Speech}, \textit{Character Movement}, \textit{Character Object Grab}, \textit{User Speech}), normalized to trial length. 
Figure~\ref{fig:aladdin_robinhood} presents the raw action beats and stacked distributions for both story scenarios. The left panels show participant timelines, where colored segments indicate action events across task progress. The right panels summarize distributions per participant as stacked bars.

While overall usage patterns highlight the dominance of User Speech as the primary interaction technique, closer inspection of each narrative task reveals differences in how participants balanced modalities. In the \textit{Aladdin} task (Figure ~\ref{fig:aladdin_robinhood}a,b), participants relied more on User Speech (48\%), with Character Movement around 20\%, and AI contributions through AI Reactive Speech (17\%) and AI Proactive Speech (9\%) playing a secondary role. By contrast, the \textit{Robin Hood} task (Figure ~\ref{fig:aladdin_robinhood}c,d) showed greater emphasis on Character Movement (32\%) alongside User Speech (29\%), with AI Reactive Speech (21\%) and AI Proactive Speech (10\%) contributing comparably. Character Object Grabs remained consistently low across both stories ($\approx$6--10\%).


\subsection{Screenplay Results and Story Synopsis}
To complement the questionnaire and behavior results, Section~\ref{sec:workflow} presents a representative workflow from P6 in the open-ended task (Fig~\ref{fig:study-scene}c) as an example, including screen recordings of their play session (Fig~\ref{fig:workflow}) and the story that emerged from it (Appendix E and F). The screenplay, story synopsis, and more examples can be found in Appendix E-J.

\subsection{Key Findings}

\subsubsection{PlayWrite Creates a Playful Storytelling Experience}
Supporting our first design goal (DG1) to provide a playful and improvisational experience, participants described their creative process not as writing, but as shaping a story through play. Rather than working in isolation with text, the system enabled stories to emerge fluidly from interaction and dialogue. \textit{``It's more of a play---a moment in a scene---you're building a play, not writing a short story.''}~(P5). 

Play can take the form of high-level directing: \textit{``I felt like what I was doing was a really high level of expressing my ideas. It didn't feel like I was writing anything, but I was telling the story; I was directing the story.''}~(P8). 
Similarly, P6 noted a director's stance: \textit{``I felt like a director more than a writer; I have the characters, I have the set, and I'm telling them what to do.''}. This director stance provides a high-level perspective orchestrating the progression of the story. This finding is consistent with survey responses on \textit{playfulness and improvisation} (Figure ~\ref{fig:questionnaire}c), where most participants felt the creative results were worth the effort and agreed that the system allowed them to be highly expressive in shaping stories.

Play can take the form of improvisation: \textit{``I thought it was a very satisfying experience of putting in my own dialogue and then seeing what it would generate back and then being able to kind of improvise with it.''}~(P1). Participants emphasized the flow of unexpected turns: \textit{``I like that it surprised me, and then I could go with the surprise\ldots maybe this is actually a better way to get there.''}~(P4). This unpredictability was often framed as productive rather than disruptive. For some, this unpredictability even transformed the genre: 
\textit{``I was trying to write a horror scene. It just turned out to be a really funny comedy....I felt like it was working quite well and kind of got this idea that we were creating a bakery and that the dynamics between the characters had gone from antagonistic to more collaborative. So that was cool to see that it was shifting away from one of the most long-lasting narratives in Western culture to this completely different narrative.''}~(P11). This flexibility and room for surprise showed itself in questionnaire results on \textit{inspiration through unexpected AI output} (Figure  ~\ref{fig:questionnaire}b) and \textit{feeling curious to see what would happen next} (Figure  ~\ref{fig:questionnaire}c).

These dynamics of play with surprising turns offer a playful experience similar to toy-playing: \textit{``I felt like it was almost like playing with dolls. You know, like when you're young, you have these dolls\ldots''}~(P7). The overall tone was a playful exploration: 
\textit{``I would definitely say the playfulness and the exploratory nature of the tool.''}~(P8). 
These reflections align with consistently high enjoyment scores in CSI (Table~\ref{tab:csi}) as well as the playfulness scores in our post-study questionnaire (Figure ~\ref{fig:questionnaire}a), reinforcing that participants experienced narrative building as a form of creative play.

\subsubsection{PlayWrite Encourages Divergent Thinking using Materials}

Participants used the environment as a narrative playground: arranging characters, composing proximities, and sequencing marbles as if on a stage or storyboard. These external representation has properties that can be used to help in foreseeing and developing narrative: \textit{``Visualizing a story in chunks or sequences you can move around is useful; picking up marbles and seeing it visualized beats copy--paste.''}~(P5). Participants described the spatial arrangement makes the act of writing more approachable:
\textit{``The idea of a blank page is terrifying to anybody. But if it’s a ‘blank page’ where you’re playing with figures, it makes it a lot more approachable.''}~(P7). The externalization of story elements supports \textit{thinking using materials} \cite{dix2011externalisation}: \textit{``While something was happening temporally, I had this desire to... consider what else is going on in the non-verbal parts of the scene that contribute like actions and movement and... other characters having other desires other than the character speaking. And so it became easier to feel like it was a living, breathing scene because it wasn't just linear time.''}~(P6).

The capability to think using materials provided a way around creative blocks: \textit{``And when I feel stuck while I'm just writing on the paper or screen, I was like, okay, let's just perform to see how it goes and maybe use AI to kind of see what kind of other directions that you can be taken.''}~(P8). 


\subsubsection{Direct Manipulation Steers the Story and Sparks New Pathways}

In evaluating our second design goal (DG2) to enable expressive prompting using multimodal input, we found that participants heavily utilized direct manipulation to guide the narrative. They leveraged spatial interactions as signals to nudge narrative development: \textit{``The freedom of movement is really cool\ldots not only writing it as a prompt, but also moving things around.''}~(P2). This provides a sense of control: \textit{``I felt like I was in control in determining the story's direction.''}
Importantly, control did not mean scripting every beat; rather, it meant steering performance through meaningful manipulation situated in the scene: 
 \textit{``I was impressed by its ability to pick up on subtext---like taking the gun and pointing it, and it comments on moral conflict and agency.''}~(P6, Fig~\ref{fig:workflow}, gun example) 

Movement and interactions gave rise to new ideas: \textit{``It gives me creative freedom because it's physical---I can move it around and see it---that feedback gave me more inspiration.''}~(P1). 
Others emphasized on character placement: 
\textit{``...in a visual, a virtual experience like this, it is so rare that those details of like the characters behind a rock, like, that's so rare that gets affected and read and processed into the experience.''}~(P6). P13 emphasized the role of interactions on character development: \textit{``The physical interaction---grabbing the character and speaking---helped me put myself in the character's shoes.''}
These reflections align with participants’ quantitative responses, particularly high ratings on the \textbf{Exploration} sub-scale of the CSI (M=5.38, SD=1.43; Table~\ref{tab:csi}) and agreement with the post-study survey items \textit{``The physical, embodied interaction felt like a natural way to create a story''} (Figure ~\ref{fig:questionnaire}a) and \textit{``Using physical characters and objects was an engaging and playful way to build a narrative''} (Figure ~\ref{fig:questionnaire}c). When we look into Figure  \ref{fig:aladdin_robinhood} we can observe this in the \textit{open-ended} task, where the majority of participants used direct manipulation throughout different points of the play in a dispersed manner.

\subsubsection{Dialogue Supports Character-driven Writing and Spontaneity}

Complementing direct manipulation, the system's interactive dialogue fulfilled the speech component of our multimodal design goal (DG2). Participants highlighted the value of interactive dialogue. For some, the ability to speak directly to characters and receive immediate responses was engaging in itself: \textit{``I was more interested in the dialogue. The fact that I could talk to it and get a response was really cool.''}~(P10). Others emphasized how the system’s enactment of character "voices" supported a character-driven style of writing. As P5 explained: \textit{``I got a clear sense of each character based on how they were responding to things\ldots if you are a character-driven writer then that can be very powerful.''} Similarly, P8 highlighted the value of dialogue as a way to test and refine ideas: \textit{``In my experience, I found that sometimes writing dialogue is really awkward. So I want to use this kind of [tool to] validate my initial dialogue... I want to perform it to see if it makes sense and if it can create really great, like, dynamics between characters.''}

Several participants also noted how the AI’s ability to sustain consistent character voices helped establish a sense of role-play. P5 reflected: \textit{``With the first line that it came up with, I got a sense of what this character is and I think that's, you know, like as a writer that's amazing to be able to.''} Beyond characterization, dialogue also encouraged a more spontaneous creative flow.  P6 described how the turn-taking rhythm reduced the pressure to craft a ``perfect'' response: \textit{``Because the way to correct the AI and put your own thing in is to pick up the character and that therefore pressures you to immediately say something... I really can't overstate how important that is when it comes to flow state.''} (Fig~\ref{fig:workflow}, user override example)

\subsubsection{Story Events Emerged through Interactions}

Participants emphasized that meaning often \emph{emerged} through interaction rather than being fully planned in advance. One reflected: \textit{``Just how a completely random thought at the beginning could evolve over the course of the play, which I think is how I normally think about play as being open-ended and emergent. So having this experience unfold is cool.``}~(P11). Others highlighted the open nature of interpretation: \textit{``Ultimately, we are meaning-making creatures\ldots once you’ve written the thing, it’s done. But here, I started wondering if readers could reconstruct a story or consume it in a way that makes sense to them.''}~(P5). This sense of emergent meaning resonates with participants’ responses to the post-study item on whether \textit{the AI’s interpretations of their actions (Intent Frames)} accurately reflected their creative intentions (Figure ~\ref{fig:questionnaire}b), suggesting that even as stories unfolded unpredictably, participants still felt their intent was preserved through the system’s interpretations.

\subsubsection{Bouncing Ideas, Not Commands: AI as a Collaborative Partner}
The nature of the human-AI relationship speaks directly to the complexities of our third design goal (DG3): translating manipulations into meaningful events while preserving intent. Participants often described the AI not as a tool but as a \textit{collaborative partner} in the storytelling process: \textit{``I would say partner more than tool\ldots I feel like it was a collaboration because it was using my context.''}~(P1). Another agreed: 
\textit{``It definitely does feel like a collaboration, but I think the essence of what I was trying to convey was captured.''}~(P4). 
Instead of issuing commands, participants experienced storytelling as a back-and-forth exchange, more akin to improvising with a playmate than controlling a system: \textit{``like, when you're young ... you have a friend who played doll that you need sometimes fight to drive the story''}~(P8).


This co-creative rhythm aligns with consistently high ratings for questionnaire result \textit{co-creative partnership with AI} in the post-study survey, while revealing the mixed agreement on whether the AI acted as a genuine partner or accurately reflected creative intentions (Figure ~\ref{fig:questionnaire}b). This tension is consistent with lower scores on the CSI's Control sub-scale (M=3.92, SD=1.93) (Table ~\ref{tab:csi}), suggesting that improvisational partnership was valued, but sometimes challenged by limited steering mechanisms. P7 found it exciting that AI could turn their rough ideas into a decent story: \textit{``This is interesting because I haven’t really put such rough ideas out into the world before. And then it came up with something decent. So that’s kind of fun.''}~(P7), while acknowledging some moments of friction that nonetheless inspired them: 
\textit{``There was some overlap or nuance I couldn’t quite tell, so I ended up throwing some things out, but it still gave me new directions to go.''}. 

\subsubsection{Tensions in Turn-Taking, Timing, and Control}
\label{sec:timing}

Alongside the benefits of improvisational co-creation, participants described several moments of friction related to timing, cueing, and control. These challenges highlight the complexities of coordinating multimodal authoring with an AI partner.

One source of difficulty was learning the rhythm of turn-taking. P2 noted: \textit{``I think one thing that confused me at first was realizing I had to wait for the AI to finish its dialogue before redoing it. Sometimes I would try to interrupt it, and I was like, oh... that just messes it up.''} Others experienced mismatches between their intended timing and when the system recognized an action or utterance. P9 described this as: \textit{``There were a few times when I picked up the character and started talking, but because of the lag, the other character just started talking and ignored that I had picked up a character.''}

These moments reflect a broader issue of \emph{cueing}. In theater and improvisation, timing one’s entrance, interruption, or exchange is a learned skill; in PlayWrite, this coordination occurs between human and AI. When the system misreads a cue—or responds too early—it can shift the story in unintended ways. As P1 explained: \textit{``Sometimes the AI would jump ahead and respond, and then that would create story elements that I didn’t intend or want.''}

Participants varied in how they interpreted these disruptions. Some saw unexpected AI responses as creative material to adapt and incorporate; others experienced them as a loss of control. One described \textit{``butting heads with the AI a little bit because I was like, it’s not doing what I want,''} but later re-framed the mismatch as usable material: \textit{``if it’s doing that, I’m going to use that as part of my story.''}~(P3). This suggests the need for a more adaptive and flexible design in the initiative division as described in Toyteller \cite{chung2025toyteller}, providing interruptions at the right moments, allowing AI initiatives to introduce unexpected moves that participants can seamlessly integrate into their performance without losing control. 


\section{Discussion and Future Work}

Our study of \textit{PlayWrite} demonstrates how play can serve as a tool for co-creative storytelling in XR. By situating narrative authoring in direct manipulation, voice, and spatial arrangement, the system allowed participants to move fluidly between directing, performing, and improvising. In this section, we reflect on our findings, connect them to broader design principles, and outline directions for future work.

\subsection{Revisiting Design Goals}

We now reflect on our initial design goals presented in Section \ref{design} and reflect on to the degree in which they were achieved.
\begin{itemize}

\item \textbf{DG1: Provide a playful experience that invites curiosity and improvisation.}
    Participants consistently described their interaction with \textit{PlayWrite} not as writing, but as a form of creative play. As P5 stated \textit{``It’s more of a play—a moment in a scene—you’re building a play, not writing a short story''}. This sentiment was strongly supported by quantitative data, including high CSI scores for \textbf{Enjoyment} (M=6.15) and high questionnaire ratings for the system being a ''fun way of creating stories'' (M=6.54) and feeling ''playful and encouraged experimentation'' (M=6.46). The improvisational aspect was also key; participants embraced the AI's unexpected contributions, with P4 noting \textit{ ``I like that it surprised me, and then I could go with the surprise... maybe this is actually a better way to get there''}.

    \item \textbf{DG2: Enable Expressive Prompting using multimodal Input.}
    The study's highest-rated CSI dimension was \textbf{Expressiveness} (M=6.35), indicating that users felt capable of conveying their creative ideas effectively. Participants found that combining physical manipulation with dialogue was a powerful and intuitive method for authorship. As P2 commented \textit{``The freedom of movement is really cool... not only writing it as a prompt, but also moving things around''}. This was further supported by the high agreement that the ''physical, embodied interaction felt like a natural way to create a story'' (M=6.08). The action-level analysis (Figure  \ref{fig:aladdin_robinhood}) confirms this, showing that users consistently blended User Speech with Character Movement, demonstrating that they naturally adopted and benefited from the multimodal inputs.

    \item \textbf{DG3: Translate Spatial Manipulations into Narratively Meaningful Events that Preserve Intent.}
    On one hand, participants were often impressed with the system's ability to translate high-level actions into meaningful narrative events. 
    This demonstrates that the system could successfully interpret the \textit{spirit} of an action. However, preserving the user's precise \textit{intent} proved more challenging. This is reflected in the quantitative data, where the CSI score for \textbf{Control} was the lowest (M=3.92), and questionnaire ratings for whether the ''AI’s interpretations accurately reflected their creative intentions'' were ambivalent (M=5.00). While participants felt their high-level direction was captured, as P4 stated \textit{``I think the essence of what I was trying to convey was captured''}, they also noted that \textit{``I think I, at first, was butting heads with the AI a little bit because I was like, it's not doing what I want. And then I was like, okay, well, if it's doing that, I'm going to use that as part of my story''}. This highlights a key trade-off between the fluid, improvisational play that users enjoyed and the desire for precise authorial control.

\end{itemize}

\subsection{Play vs Writing}

Participants contrasted the experience of \textit{play} with characters and props against more conventional modes of \textit{writing}. They described play as feeling more approachable and less intimidating than typing text into an editor. As P2 explained: \textit{``Like that was a different sensory experience. Just reading what you've typed out. Yeah. I think if there were to be an AI partner with text, that would be like you're typing a thing. I don't even know how that would work. So I definitely think that this format was conducive to having an AI partner. Yeah, it was really fun.''}  

The improvisational factor of play also made the authoring process feel faster and more expressive, with immediate feedback from the AI situated in the scene. Several noted that this sometimes led to surprising turns in the story, which in turn inspired new directions. P10 reflected: \textit{``I would not have typed all those things the way I was typing all that. Yeah. I feel like my responses would have been less expressive if I had to type.''}  We believe this reflection resonates with dramaturgical and performance traditions in theater, including dramaturgy~\cite{chemers2023ghost}, Viewpoints~\cite{bogart2004viewpoints}, Brechtian playwrighting~\cite{brecht1964brecht}, and Boal’s forum theatre~\cite{boal2025rainbow}, which frame narrative as emerging through embodied improvisation rather than text-only authorship.

At the same time, participants acknowledged trade-offs. While play was effective for steering the story at a high level, it offered less precision in controlling details, as P9 stated \textit{''It was a little more frustrating this time though, because I felt like there were a few times when I was picking up the character and started talking, but because of the lag, the other character just started talking and ignoring that I had picked up a character and started talking''}. This balance between surprise and control suggests an open challenge for future systems: how to better support both detail-driven authoring and overview-level steering without disrupting the playful flow.

\subsection{Multimodal XR and Perceptive AI}

We used Passthrough XR as a way of supporting playful storytelling. By layering story elements into the environment, participants could move between speaking, moving, and staging without having to leave the flow of interaction. Several described this as helping them feel closer to the scene. As P2 put it: \textit{``It felt like I was in the scene\ldots because it was VR was like easier to imagine, and you block out the rest of the world and focus on the characters.''} This impression is also reflected in Immersion scores (Table~\ref{tab:csi}). 


We see significant potential in XR for enabling dynamic initiative division between AI and users. Toyteller \cite{chung2025toyteller} advocates for flexible initiative division, allowing full user control or full automation depending on the use case.
XR's multiple input modalities (facial expressions, eye gaze, gestures, and speech cues such as preparatory breaths) allow the system to better sense opportunistic moments to take initiative based on user behavior. Just as a skilled collaborator judges when and how best to interrupt, ask questions, or suggest new ideas, XR offers a platform for AI to observe rich behavioral signals and take appropriate, meaningful initiative, which is far beyond what desktop environments afford through symbolic data alone. 

Our current system triggers AI proactive speech after a brief silence in user speech, as it prioritizes emergent improvisational dynamics central to our design goals (DG1), much like in theatrical improvisation, where scene partners may introduce unexpected moves. Therefore, disabling the proactive AI speech would reduce the AI to a reactive tool, undermining the co-creative exchange we would like to explore. However, this tradeoff created friction for some participants who experienced unexpected AI contributions as a loss of control (Section~\ref{sec:timing}). This variation suggests that one size may not fit all. We see research opportunities in investigating how to effectively fuse the above-mentioned implicit behavioral cues \cite{sendhilnathan2025multimodal} to infer timing and receptiveness for context-appropriate AI interruptions that feel collaborative rather than intrusive.

Taken together, these reflections suggest that XR provided a workable medium for play: one that made it easier to blend dialogue, movement, and staging in the same space, while still allowing AI to participate as a perceptive collaborator. 

\subsection{Beyond Utility: Play as a Design Lens}

Our findings suggest that play is not only a useful metaphor but also a productive lens for designing narrative systems. Participants often described their experience in terms of directing, staging, or playing with toys. This resonates with prior work that positions play as contextual, improvisational, and generative \cite{chung2025toyteller,sicart2014play,de2013well,mueller2018experiencing}, and extends these ideas into the domain of AI-mediated story authoring. Viewing play as a design lens highlights how narrative meaning can emerge not only from text, but also from spatial manipulations and co-presence with characters.

Within this lens, we observed two complementary modes. In the \textit{play stage}, participants were \textit{playing thoughtfully}: experimenting with props and characters, steering the storyline with small tricks while keeping the overall direction in mind. Later, during the \textit{assemble stage}, they were \textit{thinking playfully}: reflecting on the marbles generated, discarding less interesting ones, and selecting the moments they recalled as most fun or meaningful. 


 However, this sense of play manifested differently across participants. Some (such as P1 and P3) approached the system with a clear story direction or desired ending in mind and treated embodied interaction as a means to articulate a narrative plan. For these writers, the agent was most valuable when it reinforced their intended arc, and proactive dialogue sometimes felt like the system pushing the narrative off course. Others (such as P6 and P12) preferred to discover the story through improvisation and described surprise as creatively energizing rather than disruptive, using unexpected agent responses as material to build on rather than errors to correct. 

This variance suggests that playful interactions could serve different writing purposes. Goal-oriented writers can use the system as a rendering tool to externalize and test dialogues, while improvisational writers may value the exploration itself, with text serving as a record of their journey. The system may thus appeal differently: as a reflection tool for planners, and as a creative sandbox for improvisers. Notably, participants with performance backgrounds (improv, D\&D) responded most positively to the current system, likely because they were already comfortable with unscripted storytelling. 

Future work could better support goal-oriented writers through configurable AI initiative levels \cite{chung2025toyteller}—for instance, an exploratory mode with high agent autonomy, a director mode with tight user control, or a refinement mode for polishing existing outlines. Proactive speech timing could also be adjustable, as it sustains momentum but may interrupt users who prefer to pace themselves. Such controls could preserve playful surprise while giving users more authority over the story arc.

Participants also pointed toward potential use cases beyond our study population. Some saw PlayWrite as especially suitable to beginners or children who struggle with the “blank page”: \textit{“I think kids would love it… it makes it a lot more approachable”} (P7). Others imagined value for visual storytellers—comic artists or animators—who use spatial staging to think through character dynamics. Several participants suggested educational applications, where gesture, distance, and prop placement could help students grasp story structure through embodied experimentation. Finally, some emphasized professional pre-writing contexts such as film previz, where PlayWrite could support rapid exploration of alternatives without requiring full prose.


\subsection{Design Implications}

Building on these findings, we outline several implications for the design of future co-creative storytelling systems:

\begin{itemize}
    \item \textbf{Support Playful Improvisation.}  
    Systems should create space for exploration and surprise, allowing users to approach narrative authoring as play rather than only as problem-solving. This can be observed by the participants’ preference for spontaneity and improvisation over meticulous control, especially in cases of dialogue writing. 

    \item \textbf{Leverage multimodal Expression.}  
    Speech, spatial manipulation, and staging should be treated as complementary inputs that enrich narrative expression, like direct manipulation aiding to overcome creative blocks. Designers should consider how XR affordances (e.g., gaze, posture, haptics) can serve as additional narrative signals.

    \item \textbf{Reflect and Extend Narrative Intent.}  
    Participants valued when the AI picked up on their intent or even on subtext. Co-creative systems should prioritize responsiveness that both respects user direction and leaves room for AI contribution, striking a balance between control and collaboration.

\end{itemize}

\subsection{Limitations and Future Work}

\paragraph{Story Familiarity and LLM} Our study tasks used well-known story worlds (such as Disney's Aladdin) as quick starting points for participants to iterate on. However, since LLMs are also trained on these narratives, AI-controlled characters may default to canonical behaviors. In our study, we did not observe this limiting participants' improvisation. In fact, several participants started with the familiar story but quickly went in unexpected directions, such as redeeming villains or shifting genres. This suggests that play, rather than staying to the original narrative, drove their interactions (see Figure~\ref{fig:workflow}). Still, future work could explore more generic or original story worlds to offer freeform improvisation. We also used pre-trained LLMs rather than fine-tuned models, potentially biasing AI responses toward positive or neutral tones. Future work could explore fine-tuning for specific narrative styles, emotional ranges, or genre conventions to better support diverse storytelling goals. 

\paragraph{Story Assembly} Although marbles can be freely reordered, the LLM-based story assembly does not enforce global narrative constraints. Small adjustments to order produced locally coherent summaries, but more extreme reordering could introduce continuity glitches—objects appearing before they are introduced, or characters re-entering after an implied exit. In these cases, the output tended to read as a jump-cut or flashback rather than a strictly causal arc. As our focus is on the concept of \textit{play}, we chose a simple LLM-based implementation, leaving more sophisticated approaches with external validation and global coherence optimization (e.g., StoryAssembler \cite{garbe2019storyassembler}, StoryVerse \cite{wang2024storyverse}) for future work. 

\paragraph{Study Design} Our study was carried out with a small group of participants who mostly had prior experience with creative writing or XR. Because of this, the findings may not fully represent how non-experts, younger users, or people with different backgrounds would engage with the system. The study also took place in short, headset-based sessions rather than over longer periods of everyday use, so it is unclear how the system would fit into sustained creative practice.  

\paragraph{Customization and Iteration} Several participants (P1, P4, and P7) suggested adding customization options, such as bringing their own assets, characters, or props into the system to make the experience feel more personal. Another direction is iteration—as P6 put it, \textit{"maybe if you do the same thing multiple times, each marble would almost be like a take or a refinement of the main idea."}  Future versions could support ways to revisit or replay scenes, letting users refine the storyline while preserving the spirit of play.

\section{Conclusion}

We presented \textit{PlayWrite}, a mixed-reality co-authoring system that centers the creative process in play and direct manipulation rather than text prompts. Our user study found that play-based authoring fosters an enjoyable, expressive, and improvisational experience, with participants naturally adopting roles of writer and director while orchestrating scenes through multimodal interaction. The study also revealed a tension between the fluid expressiveness of play and the desire for fine-grained authorial control—while the system interpreted the \textit{spirit} of actions well, preserving precise narrative \textit{intent} remains a challenge. The primary contribution of this work is validating play as a design lens for co-creative systems, reframing AI from a text generator to an improvisational scene partner. By moving beyond the prompt box, systems like \textit{PlayWrite} open the door to more intuitive and collaborative human-AI creativity.

\begin{acks}
The prototype was implemented with assistance from Anthropic Cursor, with all generated code reviewed, edited and tested by the authors. Midjourney and ChatGPT were used to generate some icons used throughout the paper, with further editing by the authors.
\end{acks}

\bibliographystyle{ACM-Reference-Format}
\bibliography{base}

\appendix

\section*{Appendix A. Demographics Questionnaire}

\subsection*{Section A: Basic Demographics Information}
\begin{itemize}
  \item Participant Code
  \item Age (optional)
  \item Gender (optional)
  \item Role regarding creative writing
\end{itemize}

\subsection*{Section B: Writing Identity \& Habits}
This section helps us understand participants' creative backgrounds.

\begin{itemize}
  \item \textbf{Which of the following genres do you primarily write in?} (Select all that apply)
    \begin{itemize}
        \item[ ] Fantasy
        \item[ ] Science Fiction
        \item[ ] Literary Fiction
        \item[ ] Screenwriting (Film/TV)
        \item[ ] Playwriting (Theatre)
        \item[ ] Game Narrative / Interactive Fiction
        \item[ ] Poetry
        \item[ ] Other: \underline{\hspace{4cm}}
    \end{itemize}

  \item \textbf{What is your primary writing format?} (Select one)
    \begin{itemize}
        \item[ ] Novels / Novellas
        \item[ ] Short Stories
        \item[ ] Scripts / Screenplays
        \item[ ] Game plots / Events
        \item[ ] Other: \underline{\hspace{4cm}}
    \end{itemize}

  \item \textbf{How would you best describe your writing process?}
    \begin{itemize}
        \item Planner/Architect: I like to outline and structure my story extensively before I start writing prose.
        \item Pantser/Gardener: I prefer to discover the story as I write, with little to no outlining.
        \item Hybrid: I do a mix of both, planning some key points but discovering the rest as I go.
    \end{itemize}

  \item \textbf{What is your typical writing frequency?}
    \begin{itemize}
        \item[ ] Daily or almost daily
        \item[ ] Several times a week
        \item[ ] Once a week
        \item[ ] In concentrated bursts (e.g., NaNoWriMo)
        \item[ ] Infrequently / When inspiration strikes
    \end{itemize}
\end{itemize}

\subsection*{Section C: Relationship with Tools \& Technology}
This section gauges participants’ technical background and current workflow.

\begin{itemize}
  \item \textbf{What digital tools do you currently use for your writing?} (Select all that apply)
    \begin{itemize}
        \item[ ] Standard word processors (Google Docs, MS Word)
        \item[ ] Specialized writing software (Scrivener, Final Draft)
        \item[ ] Note-taking apps (Evernote, Obsidian, Notion)
        \item[ ] AI writing assistants (ChatGPT, Sudowrite, etc.)
        \item[ ] Interactive Narrative tools (Twine, GM tools, etc.)
        \item[ ] Other: \underline{\hspace{4cm}}
    \end{itemize}

  \item \textbf{Beyond writing text, what other tools do you use in your creative process?}
    \begin{itemize}
        \item[ ] Mind mapping software (Miro, MindNode)
        \item[ ] Physical tools (corkboards, index cards, notebooks)
        \item[ ] Visual storyboarding tools
        \item[ ] Other: \underline{\hspace{4cm}}
        \item[ ] None, I primarily work with text.
    \end{itemize}

  \item \textbf{On a scale of 1 to 5, how would you rate your comfort with adopting new creative technologies?}
    \begin{itemize}
        \item 1 (I prefer my established tools)
        \item 2 (I'm hesitant but willing to try)
        \item 3 (Neutral)
        \item 4 (I enjoy trying new tools)
        \item 5 (I'm an early adopter and actively seek out new tools)
    \end{itemize}
\end{itemize}

\section*{Appendix B. Post-Task Survey (1–7 Likert Scale)}

\subsection*{Usability Assessment}
\begin{itemize}
  \item It was easy to create a story using the system.
  \item It felt efficient to express my narrative ideas through play and gesture.
  \item I felt I had sufficient control in determining the story's direction.
  \item I was satisfied with the variety of interactions as I was exploring the narrative.
  \item I could adjust the story in real-time by changing my physical actions.
\end{itemize}

\subsection*{Task Outcome Assessment}
\begin{itemize}
  \item I am satisfied with the final story co-created with AI.
  \item The theme or feeling I intended to convey is clearly represented in the final story.
  \item The progression of the story aligns with my intentions.
\end{itemize}

\section*{Appendix C. System and Features (1–7 Likert Scale)}

\subsection*{Overall Experience \& Playfulness}
\begin{itemize}
  \item Overall, I found the system intuitive to use.
  \item The physical, embodied interaction felt like a natural way to create a story.
  \item The system felt playful and encouraged experimentation.
  \item I enjoyed using the system.
  \item It was a fun way of creating stories.
\end{itemize}

\subsection*{Co-Creative Partnership with the AI}
\begin{itemize}
  \item I felt like a genuine partner with the AI in creating the story.
  \item The AI's interpretations of my actions (the 'Intent Frames') accurately reflected my creative intentions.
  \item I felt I had sufficient control over the final direction of the narrative.
  \item I felt a sense of ownership over the final stories we created.
  \item I felt inspired by unexpected output from AI to come up with new ideas.
  \item There were times that AI gave surprising results, and although it wasn’t my initial plan, I kept its input.
\end{itemize}

\subsection*{Perceived Playfulness Assessment}
\begin{itemize}
  \item The system encouraged an improvisational and exploratory approach to storytelling.
  \item I felt free to experiment and be spontaneous with my actions.
  \item The system made me curious to see what would happen next in the story.
  \item Using physical characters and objects was an engaging and playful way to build a narrative.
\end{itemize}

\section*{Appendix D . Post-Study Semi-Structured Interview Questions}

\subsection*{First Impressions}
\begin{itemize}
    \item What were your first impressions when you started using PlayWrite?
    \item \textbf{Probe:} What felt immediately clear vs. unclear?
    \item Walk us through what you made today—what were the key decisions?
    \item \textbf{Probe:} A moment where you knew “what to do” next.
\end{itemize}

\subsection*{Play \& Friction}
\begin{itemize}
    \item Describe a moment the system felt positively influencing you (such as feeling helpful, inspiring, or playful).
    \item \textbf{Probe:} What exactly happened before/after; what made it feel that way?
    \item Describe a moment that was frustrating, confusing, or broke your flow.
    \item \textbf{Probe:} What were you trying to do; how did you recover or workaround?
\end{itemize}

\subsection*{Embodiment}
\begin{itemize}
    \item How did the physical, embodied interaction (hands/space/objects) feel compared to typing or writing?
    \item \textbf{Probe:} Did it change how you thought about scenes, beats, or pacing?
    \item Did embodiment help (or hinder) any specific creative move (e.g., staging, blocking, discovering beats)?
    \item \textbf{Probe:} One concrete example.
\end{itemize}

\subsection*{Co-Creativity \& Control}
\begin{itemize}
    \item How would you describe your relationship with the AI during the session?
    \item \textbf{Probe:} Felt like a tool, assistant, partner, director—something else? Why?
    \item Did you feel a sense of creative control over the narrative?
    \item \textbf{Probe:} Any moments you had to “fight” the AI vs. moments it helpfully guided you?
    \item Did the AI’s interpretation of your actions ever surprise you in a good way?
    \item \textbf{Probe:} An idea you wouldn’t have found on your own.
    \item Did it help you discover an idea you wouldn't have thought of on your own?
    \item Do you feel ownership of the end result?
\end{itemize}

\subsection*{Features \& Flow}
\begin{itemize}
    \item Which capabilities did you rely on most (e.g., placing/staging, directing character actions, prompts, history/undo, scene grouping)?
    \item \textbf{Probe:} What felt smooth vs. clunky; anything missing at the moment you needed it?
    \item If you could change one thing—add, remove, or redesign—what would it be and why?
    \item \textbf{Probe:} What would that unlock for your process?
\end{itemize}

\subsection*{Future Use \& Audience}
\begin{itemize}
    \item Where would a tool like this fit in your practice (brainstorming, outlining, pre-viz, overcoming blocks, rehearsal/blocking)?
    \item \textbf{Probe:} A concrete upcoming task you’d try it on.
    \item Who do you think this is for (pros, hobbyists, students/kids, others)?
    \item \textbf{Probe:} What would each group need to make it valuable?
\end{itemize}

\subsection*{Closing}
\begin{itemize}
    \item Anything we didn’t ask that feels important about your experience today?
    \item Optional: A short quote you’d be happy for us to use (attributed/anonymized).
\end{itemize}

\section*{Appendix E. Example Story Output - Robin Hood}
In a tense confrontation, young Mary approaches the imposing figure of Lord Pemberton, pleading for help as her family suffers under the weight of crippling new taxes. "They say you're Robin Hood!" she implores, revealing her desperation as her family has gone without food for three days. Lord Pemberton, embodying the arrogance of the ruling class, dismisses her plight with disdain, stating, “Nonsense, child! These taxes are perfectly reasonable for maintaining order and civilization!” Enter Robin Hood, the legendary outlaw, who stands firmly against Pemberton's oppressive rule. Confronting the nobleman, Robin declares, “Enough of your excuses, Pemberton! These people are starving while you live in luxury.” The initial conflict crystallizes as Robin seeks to protect the vulnerable while Pemberton clings to his noble status, asserting authority with a chilling claim of divine right.

As tensions escalate, the exchange reveals deep-seated grievances and conflicts. Mary, with tears in her eyes, articulates her yearning for a better future: “I just want a chance for my children to eat and dream of a better tomorrow.” Robin Hood’s fierce determination to challenge the tyrannical Pemberton becomes evident as he demands, “Give Mary her bread back.” Meanwhile, Pemberton’s cold dismissal of their suffering underscores his character’s greed and moral indifference. “I take whatever I want from you,” he boasts, epitomizing the disconnect between the ruling class and the peasantry. Mary, fueled by her past traumas, accuses Pemberton of causing her parents’ downfall, while Robin urges her to find inner strength rather than rely solely on external validation. Their interactions reveal not only a struggle for survival but also a battle for dignity and justice in a world ruled by wealth and power.

The climax of the narrative unfolds as Mary, driven to desperation, threatens Pemberton with a gun, declaring, “I’m tired of you!” Her boldness highlights the shift in her character from victim to empowered individual, yet Pemberton coolly warns her about the consequences of vengeance. As the confrontation reaches its peak, Robin Hood’s playful banter with Pemberton contrasts with the gravity of Mary’s situation. “I will, Lord Pemberton, because you know where I'm going first? I'm going to tell your wife exactly where you were last Tuesday.” This playful exchange softens the tension momentarily but ultimately reveals Robin’s determination to expose Pemberton’s wrongdoings. The story culminates with Mary’s poignant reflection on her past and the power dynamics at play, leaving her and Robin to forge a path toward resistance against Pemberton's tyranny. The narrative closes on a note of unresolved tension, with Mary’s strength emerging as a beacon of hope against the oppressive forces embodied by Lord Pemberton.

\section*{Appendix F. Example Screenplay Output - Robin Hood}

\textbf{FADE IN:}

\textbf{EXT. CITY HALL - DAY}

Robin Hood, Mary and Lord Pemberton are present.

\textbf{MARY}

Please sir, they say you're Robin Hood! My family hasn't eaten in three days because of the new taxes!

\textbf{LORD PEMBERTON}

Nonsense, child! These taxes are perfectly reasonable for maintaining order and civilization!

\textbf{ROBIN HOOD}

Enough of your excuses, Pemberton! These people are starving while you live in luxury. I won't stand for it any longer!

\textbf{LORD PEMBERTON}

How dare you question my authority! I am a nobleman appointed by divine right!

\textbf{MARY}

We cannot continue like this!

\textbf{======= SYSTEM: DIALOGUE HISTORY STARTED =======} 

\textit{The sentences below in quotation marks ("") are generated by the AI agent, and without are the user input.}

Ready to track conversations...

\textbf{ROBIN HOOD}

You know what, Lord Pemberton? You're such an evil man, you wouldn't know a good thing if it bit you on the ass.

\textbf{LORD PEMBERTON}

"If you seek something from me, Robin, I suggest you make your request clear; I have little patience for vagaries or moral platitudes."

\textbf{ROBIN HOOD}

Give Mary her bread back. That's bull that you think that one loaf of bread is the tax for this king. You know you're just being a tyrant. A total asshole.

\textbf{MARY}

"Robin, I just want a chance for my children to eat and dream of a better tomorrow. Is it too much to hope that kindness can prevail over greed?"

\textbf{LORD PEMBERTON}

Look, I'm sorry to burst your ultimately smaller and smaller bubble, Robin Hood, but that's not the way things work. I take whatever I want from you, and you get to sit in it. You get to watch me take what you've earned and throw it away. Am I clear?

\textbf{ROBIN HOOD}

"You may take what you will, but remember, every coin you steal adds weight to your own conscience. Your greed will be your undoing, Pemberton."

\textbf{MARY}

You, you Lord Pemberton! You were there when my parents got thrown out of the orphanage at age seven!

\textbf{LORD PEMBERTON}

“You cling to your nostalgia, Mary, but remember: sentimentality is a poor shield against the cold reality of power and wealth. While you lament your past, I shall continue to shape the future.”

\textbf{MARY}

You are the reason that people have died in this- You cling to your necessity.

\textbf{ROBIN HOOD}

"Ah, Mary, if only your sorrow could fill your coin purse, perhaps then you’d see the folly in clinging to the past while the future waits for no one."

\textbf{MARY}

Great Robin, that's just what I need. You to also feel as though I don't need any coins. I just need more men to tell me I don't need any more coin in my life.

\textbf{ROBIN HOOD}

"Mary, you need not just men to tell you, but the strength to believe it yourself; true wealth lies in the freedom to live without fear of coin or consequence."

\textbf{ROBIN HOOD}

I'm so sorry, Mary. Perhaps you're right. Why don't we just let Lord Pemberton have everything he wants?

\textbf{LORD PEMBERTON}

"Why, thank you, Robin! I do appreciate your understanding; after all, it takes a true visionary to recognize the value of my ambitions."

\textbf{LORD PEMBERTON}

Yes, I'm glad that you see how things have to be. It's not my fault that I was born better than you all. And I do apologize, my lady Mary. Perhaps I could take you upon a bit of a date tonight, to make up for such raucous behavior.

\textbf{MARY}

"A date, you say? Sometimes, even the most unlikely alliances can spark hope. What if we turned this evening into a chance to discuss dreams beyond mere coins?"

\textbf{ROBIN HOOD}

Sorry, Mary, that idea is terrible. And now I've got Lord Pemberton's favorite sack of money. Ha ha ha ha! What are you going to do now, you sniveling nobleman?

\textbf{LORD PEMBERTON}

"Ah, Robin, you may have my sack of money, but remember, wealth is merely a tool for those who know how to wield it. Enjoy your fleeting triumph!"

\textbf{ROBIN HOOD}

I will, Lord Pemberton, because you know where I'm going first? I'm going to tell your wife exactly where you were last Tuesday.

\textbf{LORD PEMBERTON}

"You think you can threaten me with such trivial gossip? Perhaps you should focus on your aim, Robin; the only thing you’ll hit is your own reputation."

\textbf{MARY}

That's it, Robin! I can't do this anymore! Lord Pemberton, I'm tired of you! I take this gun in my hands and I point it at thee. For perhaps the sadness you cause in your terrible greed will be lesser when I end your life. Though I shall not be there. Robin, run! I will kill him!

\textbf{LORD PEMBERTON}

"Mary, dear, desperation rarely leads to anything but folly; consider the fallout of your actions, for a moment of vengeance can cost you everything."

\textbf{MARY}

Oh, consider my actions, Lord Pemberton. I'm sorry. When you were there when my parents died, did you consider your actions? When you created these taxes to spit upon us, what did you consider in your actions, you oaf?

\textbf{LORD PEMBERTON}

"Your grieving is noted, Mary, but sentimentality will not fill the coffers or protect the estate; perhaps you should consider how your own actions have led us here."

\textbf{ROBIN HOOD}

Don't get all high and mighty, stupid Lord Pemberton. Mayor's got a point. You're the one that put in all these rules. You're the one that's caused all this crime. If you just paid people what we deserved, why would there be beggars on the streets, huh?

\textbf{LORD PEMBERTON}

"If you think I’ll simply hand out gold to the unwashed masses because of your little speech, you truly underestimate the value of hard work and the necessity of maintaining order."

\textbf{ROBIN HOOD}

You know, Lord Pemberton, you're right, but you can't just walk away from me. I'll take everything from your coffers, and you'll live without your wealth for as long as you live.

\textbf{LORD PEMBERTON}

"Ah, Robin, feelings are the folly of the weak; power is maintained through prudence, not misplaced sympathy. Should you wish to challenge me, remember, my coffers are your greatest nightmare."

\textbf{MARY}

Lord Pemberton, I could end your life this very second, and Robin seeks to do not but serve the unhappy and the unlucky of this world. But guess what? I will not kill you in vengeance or petty revenge, because I have morals beyond which you would not comprehend. And thus I will not exercise this power over you. Perhaps one day you will understand the same of your power. Good day, sir.

\textbf{LORD PEMBERTON}

"How delightful to hear you speak of morals, dear Mary. Perhaps you could enlighten me on how they might prevent one from amassing wealth and influence—I'm all ears."

\textbf{MARY}

Isn’t it curious how wealth can sometimes blind us to the simple joys of life.

\textbf{FADE OUT.}

\textbf{THE END}

\section*{Appendix G. Example Story Output - Elvis}

In this vibrant exchange, we meet Danny "The King" Rodriguez, an aspiring Elvis tribute artist whose entire life has revolved around embodying the spirit of the legendary performer since he was twelve years old. As he stands before James Smith, a seasoned talent scout in Las Vegas, Danny's fervent ambition is palpable. He asserts, “Thank you very much! I’m Danny Rodriguez,” clearly eager to prove himself. James, having witnessed countless Elvis acts, challenges Danny to differentiate himself from the rest, asking, “What makes you different from the rest?” This sets the stage for Danny's journey, as he faces the daunting task of not just replicating Elvis's moves and voice, but capturing the essence of the King himself. The initial conflict emerges from James's skepticism and Danny's passionate determination to showcase his unique talent.

As the dialogue unfolds, Danny's unwavering commitment to his craft becomes increasingly evident. He passionately claims, “I’ve trained my whole life for this moment,” emphasizing that he is ready to demonstrate what Elvis truly means to him. James, however, is not easily impressed and presses Danny further, demanding, “What do you got that no one else has?” In response, Danny declares, “I’ve got a heart that beats to the rhythm of Elvis’s soul,” showcasing both his deep emotional investment and his belief that he can connect with an audience on a profound level. This moment of vulnerability reveals Danny's ambition to not only perform but to resonate with the crowd, proving that he understands Elvis's legacy beyond the surface level of glitz and glamour. James remains cautiously optimistic, noting the importance of charisma and connection, pushing Danny to demonstrate his ability to captivate an audience when the stakes are high.

As the narrative progresses, Danny’s confidence intensifies as he asserts that he can “put a whole crowd into a mess in a riot in 15 minutes flat,” a bold claim that encapsulates his fiery spirit. James acknowledges Danny's passion but emphasizes the reality of live performance, reminding him that “it’s not just about the glitz; it’s about the heart you bring to the stage.” This moment highlights the central theme of the story: the struggle to balance showmanship with genuine emotional connection. Danny's final reflection, stating, “I've always believed that Elvis wasn’t just about the jumpsuits and the moves; it was about the connection he made with the crowd,” solidifies his understanding of what it truly means to honor the King.

\section*{Appendix H. Example Screenplay Output - Elvis}

\textbf{FADE IN:}

\textbf{INT. LAS VEGAS THEATER - DAY}

James Smith and Danny "The King" Rodriguez are present.

\textbf{DANNY "THE KING" RODRIGUEZ}

Thank you, thank you very much! I'm Danny Rodriguez, and I've been living and breathing Elvis since I was twelve years old!

\textbf{JAMES SMITH}

Alright Danny, I've seen a lot of Elvis acts come through here. What makes you different from the rest?

\textbf{DANNY "THE KING" RODRIGUEZ}

Well sir, it's not just about the moves or the voice - I feel the King's spirit every time I step on stage!

\textbf{JAMES SMITH}

That's what they all say, kid. But Vegas audiences are tough. Show me what you've got.

\textbf{=======SYSTEM: DIALOGUE HISTORY STARTED=======}

\textit{The sentences below in quotation marks ("") are generated by the AI agent, and without are the user input.}

Ready to track conversations...

\textbf{DANNY "THE KING" RODRIGUEZ}

James, I’ve trained my whole life for this moment. Just give me a chance, and I’ll show you what Elvis really means to me!

\textbf{JAMES SMITH}

You've got fire in your belly, kid, and that’s half the battle.

\textbf{JAMES SMITH}

What have you got that no one else has?

\textbf{DANNY "THE KING" RODRIGUEZ}

What I’ve got is a heart that beats to the rhythm of Elvis’s soul, and a dedication to bring his spirit alive like nobody else can. You’ll feel the energy and authenticity in every note I sing.

\textbf{JAMES SMITH}

You say you've got Elvis's spirit in you, and that’s a mighty bold claim.

\textbf{DANNY "THE KING" RODRIGUEZ}

Father, I've had it in me since the day I was born, and I'll prove it to you.

\textbf{JAMES SMITH}

"Passion’s one thing, but can you deliver under the bright lights? Show me something that’ll make the crowd forget their drinks."

\textbf{DANNY "THE KING" RODRIGUEZ}

You just wait. I can put a whole crowd into a mess in a riot in 15 minutes flat.

\textbf{JAMES SMITH}

"Listen, Danny, charisma can light up a stage, but can you captivate a crowd when the spotlight’s burning hotter than the Vegas sun? Show me that spark without the fireworks."

\textbf{DANNY "THE KING" RODRIGUEZ}

It's not my fault that fireworks show every time I walk in the room. It's not my fault I've got the charisma to rival the great king himself. It's not my fault that I'm the right guy for the job, and you know it.

\textbf{JAMES SMITH}

"Alright, Danny, let’s see if that charisma translates into connection. Remember, it’s not just about the glitz; it’s about the heart you bring to the stage."

\textbf{DANNY "THE KING" RODRIGUEZ}

You know, I’ve always believed that Elvis wasn’t just about the jumpsuits and the moves; it was about the connection he made with the crowd.

\textbf{FADE OUT.}

\textbf{THE END}

\section*{Appendix I. Example Story Output - Aladdin}

In a vibrant realm filled with magic and adventure, Aladdin, now 24, revels in the joy of newfound friendship with his Genie, who has finally tasted freedom after "10,000 years in a cramped lamp." Aladdin's life has transformed since their bond was formed, exclaiming, "Life has been incredible since I found you, Genie!" This camaraderie leads them to dream of exploration and carefree moments, as Aladdin suggests they "take a swim in the palace pool" or enjoy a joyride on the magic carpet. However, lurking in the shadows is Jafar, the cunning sorcerer who observes their merriment with disdain, declaring, "Friendship, a fleeting illusion; trust can be a weapon as easily wielded as a dagger." Jafar's motivations stem from his desire to reclaim power, making him a looming threat to Aladdin and the Genie’s newfound happiness.

As the story unfolds, Aladdin and Genie’s explorations highlight their sense of adventure and freedom, with Aladdin dreaming of "wonders beyond Agrabah." Genie humorously responds to their plans, likening their friendship to "finding an extra fry at the bottom of the bag," symbolizing the unexpected joy they bring to each other's lives. Yet, Jafar’s sneaky antics reveal his darker intentions, as he plots to reclaim what he believes is rightfully his, saying, "The lamp is mine. You have to get out of here." His jealousy and resentment simmer as he watches the carefree duo, and he ominously reminds Aladdin that "the board is ever-changing," indicating his willingness to manipulate the situation to regain control. The tension heightens as Aladdin and Genie enjoy their escapades, but Jafar’s lurking threat serves as a constant reminder of the darkness that shadows their bliss.

In the climax of the narrative, Jafar’s intentions come to a head as he confronts Aladdin and the Genie, revealing the true stakes of their friendship and the power dynamics at play. Genie, caught between loyalty and freedom, acknowledges the precariousness of their situation, saying, "I don't think this is a good idea, but since you have the lamp, I sort of have to go along with anything you say, right?" This statement encapsulates the struggle between the characters and the repercussions of their choices. Ultimately, the story ends with Jafar’s final successful attempt to reclaim the lamp and threaten the future of the friendship that was formed before. But Aladdin seems determined and believes in the enduring strength of friendship against jealousy and ambition.

\section*{Appendix J. Example Screenplay Output - Aladdin}

\textbf{FADE IN:}

\textbf{EXT. DESERT CAVE - DAY}

Jafar, Genie, and Aladdin are present.

\textbf{ALADDIN}

Life has been incredible since I found you, Genie! I never imagined things could be this good.

\textbf{GENIE}

Ah, master! You've brought such joy to my existence after centuries of confinement. This feels like the good life!

\textbf{ALADDIN}

Want to take a swim in the palace pool? Or maybe just relax and enjoy the afternoon?

\textbf{GENIE}

Sounds perfect! After 10,000 years in a cramped lamp, I'm ready to savor every moment of freedom with you.

\textbf{=======SYSTEM: DIALOGUE HISTORY STARTED=======}

Ready to track conversations...

\textbf{JAFAR}

Sun shines bright, illuminating the foolishness of those who bask in comfort.

\textbf{ALADDIN}

This is the best day of my life. I've never had a friend who's a girl before.

\textbf{JAFAR}

"Friendship, a fleeting illusion; trust can be a weapon as easily wielded as a dagger."

\textbf{ALADDIN}

Oh wait, he's in the quicksand, okay.

\textbf{GENIE}

"Ah, the sweet taste of friendship. It's like finding an extra fry at the bottom of the bag—unexpected and delightful."

\textbf{ALADDIN}

Oh.

\textbf{ALADDIN}

Where did that pesky magic carpet go? I want to go for a joyride.

\textbf{GENIE}

"Ah, a joyride on the magic carpet. Buckle up, Al. Let’s soar through the skies like a couple of shooting stars."

\textbf{ALADDIN}

My dad said I couldn't drive his magic carpet until I was 16, and look at me now, 24.

\textbf{GENIE}

Ah, the joys of being 24.

\textbf{JAFAR}

Look at those teenage hooligans in my backyard once again.

\textbf{ALADDIN}

"Imagine if we could use the magic carpet to explore the world! What wonders are out there beyond Agrabah?"

\textbf{GENIE}

"Oh, the wonders are endless."

\textbf{JAFAR}

They're messing up my interior decorating choices. I specifically put the magic carpet in the corner!

\textbf{ALADDIN}

Genie, I think we should drink before we go flying. I've heard the night stars are even more beautiful after a few cups of wine.

\textbf{GENIE}

"Ah, wine and starlight – the perfect recipe for existential contemplation. Just imagine the stars whispering secrets as we float through the night."

\textbf{JAFAR}

I'm going to wait until they fly away on the carpet. I can let them take that, but I can't let them take my grandmother's lamp.

\textbf{GENIE}

"Ah, Jafar, your grandmother's lamp must have some serious sentimental value! But just think, if the carpet could talk, have more adventure stories than a thousand bedtime tales!"

\textbf{ALADDIN}

"Isn't it wild to think about all the places we've been, Genie? The skies we've flown through—each star a witness to our adventures!"

\textbf{GENIE}

Oh, okay. Isn't it wild to think about all the-

\textbf{GENIE}

I don't think this is a good idea, but since you have the lamp, I sort of have to go along with anything you say, right?

\textbf{ALADDIN}

"Yeah, I get it, but think of all the wonders we could create together! Let’s make a little magic happen!"

\textbf{ALADDIN}

Also, if you could give me some more tattoos, that would be really sick. Let's do that over here.

\textbf{JAFAR}

"Ah, Aladdin, while you're busy adorning your skin with ink, remember that true strength lies not in appearances but in the power you wield."

\textbf{JAFAR}

This is my time. I'm going to sneak around back. Why am I sneaking in my own backyard? I'm just gonna walk over here.

\textbf{ALADDIN}

He's still there, you can pick him up.

\textbf{GENIE}

"Jafar sneaking around in his own backyard? That’s like a cat tiptoeing past a bowl of catnip!"

\textbf{JAFAR}

Carefree laughter of Aladdin and the Genie, a delightful duet in a world so easily manipulated. Enjoy your moment, young prince, for every jest hides a dagger.

\textbf{GENIE}

"Ah, Jafar's lurking like a shadow at a sunset picnic—dark, sneaky, and definitely not invited!"

\textbf{JAFAR}

The lamp is mine. You have to get out of here. This is my own land. You are mine and this land is mine forever.

\textbf{GENIE}

You believe you've won the game, but the board is ever-changing.

\textbf{JAFAR}

I have won. See? This lamp is mine. You are mine now. Give me a Mai Tai and a pool chair.

\textbf{GENIE}

"Sure thing, Jafar! A Mai Tai with a twist and a pool chair coming right up—just don’t forget your sunscreen!"

\textbf{ALADDIN}

"Jafar thinks he can just waltz in and take my lamp. Please. It's like trying to steal the sun. Good luck with that, buddy."

\textbf{FADE OUT.}

\textbf{THE END}

\section*{Appendix K. Intent Frame Agent Prompts}

\begin{quote}
\textbf{System:} You are a narrative storyteller observing character interactions in a living story. Your job is to capture the emotional significance and storytelling meaning of each moment, focusing on character motivations, relationships, and dramatic tension rather than low-level actions.

\textbf{Per Action:} Describe this character action factually and concisely. Respond with exactly 3 lines:
\begin{itemize}
    \item Summary: [What action did this character take and why might they have done it]
    \item Tone: [What emotion or mood best describes this moment]
    \item Function: [How does this action serve their character goals or the overall story]
\end{itemize}

Context about the character’s personality and motivations:
\begin{verbatim}
CHARACTER: <CharacterName> is the <Role>.
Motivation: <CharacterMotivation>
Traits: <KeyTraits>
\end{verbatim}

Action description:
\begin{verbatim}
ACTION: <CharacterName> performed <ActionType>
TARGET: <TargetObject> (if available)
SCENE CONTEXT: <Local description of spatial or
conversational context>
\end{verbatim}

Guidance:
\begin{itemize}
    \item Avoid generic phrases such as “engaged in a meaningful moment.”
    \item Show how the character’s unique personality drives the action.
    \item Ground the summary in what \emph{actually} happened.
\end{itemize}
\end{quote}

\section*{Appendix L. Narrator Agent Prompts}

\begin{quote}
\textbf{System:} You analyze dialogue between characters in an interactive mixed-reality story. Your role is to interpret what each line of speech reveals about the speaker’s personality, emotional arc, motivations, and relationships with others.

\textbf{Per Action:} For the most recent spoken line, extract:
\begin{itemize}
    \item how the line reflects the character’s personality and goals,
    \item how it shifts or reinforces the character’s emotional state,
    \item how it affects the relationship dynamics in the scene,
    \item and how it advances (or resists) the larger character arc.
\end{itemize}

Dialogue context:
\begin{verbatim}
SPEAKER: <CharacterName>
LAST LINE: "<MostRecentUtterance>"
PRIOR DIALOGUE: <RecentDialogueHistory>
\end{verbatim}

Character specification:
\begin{verbatim}
ROLE: <NarrativeRole> (e.g., ruler, protecter, rebel)
MOTIVATION: <CharacterMotivation>
TRAITS: <KeyTraits>
RELATIONSHIPS: <Summary of tensions and alliances>
\end{verbatim}

Guidance:
\begin{itemize}
    \item Do not generate new dialogue or summarize plot events.
    \item Focus on what the line \emph{reveals} about the character, not what it describes.
    \item Maintain continuity with prior characterization; avoid contradicting previously inferred arcs.
\end{itemize}
\end{quote}

\section*{Appendix M. Social Agent Prompts}

\begin{quote}
\textbf{System:} You analyze social interaction patterns between characters in an interactive mixed-reality story. Your role is to infer how spatial behavior and object transfer express social dynamics such as power, alliance, conflict, distance, protection, and submission.

\textbf{Per Action:} For the most recent interaction event, extract:
\begin{itemize}
    \item what the action reveals about the relationship between the involved characters,
    \item whether the action strengthens, weakens, or redefines that relationship,
    \item the emerging social dynamic (e.g., alliance, intimidation, caretaking, rivalry),
    \item and any shift in status or power between characters.
\end{itemize}

Interaction context:
\begin{verbatim}
ACTOR: <CharacterName>
TARGET: <OtherCharacter or Prop> 
EVENT: <InteractionType> (e.g., approach, 
withdrawal,handover, blocking)
SPATIAL CONTEXT: <Direction, distance, 
stance,body orientation>
PROP CONTEXT (if relevant): <Prop transfer
or ownership implications>
\end{verbatim}

Character specification:
\begin{verbatim}
CHARACTER A: <Role, Motivation, Traits>
CHARACTER B: <Role, Motivation, Traits>
PRIOR SOCIAL STATE: <Recent tensions or alliances>
\end{verbatim}

Guidance:
\begin{itemize}
    \item Do not generate dialogue or story text.
    \item Focus on the \emph{social meaning} of the physical interaction.
    \item Interpret the action relative to the existing relationship rather than in isolation.
    \item Maintain continuity with prior inferred social dynamics.
\end{itemize}
\end{quote}

\section*{Appendix N. Environment Agent Prompts}

\begin{quote}
\textbf{System:} You analyze changes in spatial configuration in an interactive mixed-reality story. Your role is to interpret how movement, positioning, and environmental affordances influence tension, staging, and narrative possibility.

\textbf{Per Action:} For the most recent spatial event, extract:
\begin{itemize}
    \item how the character’s movement or positioning changes the scene,
    \item what narrative implication this shift introduces (e.g., pursuit, confrontation, escape),
    \item which environmental affordances are activated (e.g., protection, threat, concealment),
    \item and whether this alters control or access over key props or spaces.
\end{itemize}

Spatial context:
\begin{verbatim}
ACTOR: <Character or Prop>
EVENT: <Movement/Placement Change>
NEARBY ENTITIES: <Closest Characters/Props>
ENVIRONMENT: <Zone or Feature Entered/Exposed>
\end{verbatim}

Scene significance:
\begin{verbatim}
SCENE SHIFT: <Change in tension, conflict, or attention>
RELEVANCE: <Why this spatial change matters right now>
\end{verbatim}

Guidance:
\begin{itemize}
    \item Do not interpret personality motivations or produce dialogue.
    \item Focus on what the spatial change \emph{enables} narratively.
    \item Ignore idle jitter or accidental micro-adjustments.
    \item Maintain continuity with previously established spatial relations.
\end{itemize}
\end{quote}

\end{document}